\documentclass[aps,twocolumn,preprintnumbers,superscriptaddress]{revtex4} 
\usepackage{amsmath,amssymb} 
\usepackage{verbatim}
\usepackage{graphicx} 
\usepackage{array} 
\usepackage{latexsym} 
\usepackage{booktabs}

\def\aver#1{{\left<{#1}\right>}}
\newcommand\be{\begin{equation}}
\newcommand\fin{\end{equation}}

\begin{document}

\title{Exact calculations of first-passage quantities on recursive networks}

\date{\today}

\author{B. Meyer}
\affiliation{Laboratoire de Physique Th\'eorique de la Mati\`ere Condens\'ee, CNRS UMR 7600, case courrier 121, Universit\'e Paris 6, 4 Place Jussieu, 75255
Paris Cedex}

\author{E. Agliari}
\affiliation{Laboratoire de Physique Th\'eorique de la Mati\`ere Condens\'ee, CNRS UMR 7600, case courrier 121, Universit\'e Paris 6, 4 Place Jussieu, 75255 Paris Cedex}
\affiliation{Dipartimento di Fisica, Universit\'a di Parma, Viale Usberti 7/A, 43100 Parma, Italy}

\author{O. B\'enichou}
\affiliation{Laboratoire de Physique Th\'eorique de la Mati\`ere Condens\'ee, CNRS UMR 7600, case courrier 121, Universit\'e Paris 6, 4 Place Jussieu, 75255
Paris Cedex}

\author{R. Voituriez}
\affiliation{Laboratoire de Physique Th\'eorique de la Mati\`ere Condens\'ee, CNRS UMR 7600, case courrier 121, Universit\'e Paris 6, 4 Place Jussieu, 75255
Paris Cedex}

\begin{abstract}
We present general methods to exactly calculate mean-first passage quantities on self-similar networks defined recursively. In particular, we calculate the mean first-passage time and  the splitting probabilities associated to a source and one or several targets; averaged quantities over a given set of sources (e.g., same-connectivity nodes) are also derived.
The exact estimate of such quantities highlights the dependency of first-passage processes with respect to the source-target distance, which has recently revealed to be a key parameter to characterize transport in complex media. 
We explicitly perform calculations for different classes of recursive networks (finitely ramified fractals, scale-free (trans)fractals, non-fractals, mixtures between fractals and non-fractals, non-decimable hierarchical graphs) of arbitrary size. Our approach unifies and significantly extends the available results in the field.
\end{abstract}

\maketitle


As they account for transport efficiency in complex media, first-passage processes \cite{Redner:2008} on fractal or complex networks have given rise to a lot of interest in the past few years. Such networks, showing a very broad spectrum of  topological  and thus
transport characteristics, have been opportunely  used to model a wide range of systems, from biology to sociology or computer science \cite{ D.Ben-Avraham:2000,Albert:2001RMP, Dorogovtsev:2008,Barratbook}. Within this outline,  the exact determination of first-passage quantities has been the topic of many works \cite{Aldous:1999,Noh:2004a,Kozak:2002PRE,Sood:2005,Agliari:2008PRE,Haynes:2008PRE,Zhang:2009fl,Tejedor:2009}, in a dialog with parallel investigations of general scalings and asymptotic results  \cite{Kahng:1989fk,Bollt:2005,Condamin:2007zl,Benichou:2008PRL}.
The global mean first-passage time (GMFPT), defined as  the mean first-passage time of a random walk starting from an arbitrary site (source) in presence of a fixed trap (target), and in particular its scaling with the number of nodes $N$, was meant to encompass the general properties of transport on fractal and transfractal networks. Yet, recent results \cite{Condamin:2007zl} have highlighted the insufficiency of this mere description and put forward the role of non-averaged quantities, associated to  a single  starting point.
For example, in the context of diffusion limited reactions in complex media,  the initial position of the reactants  has indeed been shown to be a key parameter  that can control the entire kinetics of the process \cite{Benichou:2010NatChem}. 

In that context, we present a general algebraic method to calculate exactly  first-passage quantities on self-similar networks, for a {\it given} source point. More precisely, we consider here two first-passage observables:
\begin{itemize}
 \item the mean first-passage time (MFPT) from a site $S$ to a site $T$, or to several sites $\{T_i\}$, i.e. the average time it takes a random walker that starts at $S$ to reach $T$ or any point in $\{ T_i \}$;
 \item the splitting probability, defined in presence of several targets $T_{i}$ as the probability, starting from $S$, to reach a given target $T_{i_0}$ before all the other targets.
 \end{itemize}
 Actually, these quantities both satisfy  simple but formal $N \times N$ linear systems, where $N$ is the number of sites of the network. The regime of interest is typically $N$ large and makes the explicit resolution of such linear systems out of reach.  The aim of the present work is to provide a general method, applicable to a broad class of networks, which  yields explicit and exact  formulas for the MFPT and splitting probabilities even for very large $N$. To do so, we make use of  the self-similar  properties of the networks considered and develop  a general renormalization scheme following ideas presented in \cite{Haynes:2008PRE}.
 

More precisely, we consider in this paper self-similar networks, which can be defined recursively \cite{D.Ben-Avraham:2000} by considering the result of the action of   a similarity transformation $r(\Gamma)$  that maps all points $i$ of a network $\Gamma$  onto new points $i' = r( i)$.  Considering that the transformation $r$ is homothetic of ratio $\rho<1$, a network  is called self-similar  if at generation $n$ its iteration  $\Gamma_n$ is equal to the union of $p$ replicas of $r(\Gamma_{n-1})$.  As a result, self-similar networks  can be split into a finite number of equal sub-units, those sub-units being the network at the previous generation. Explicit examples will be given throughout this paper.

An important point  is the relationship between self-similarity and fractality.  A network is called fractal if one can define a constant $d_f$ such as $N \sim R^{d_f}$, where $N$ denotes the volume (or number of nodes) and $R$ the chemical diameter. Note that whereas most examples of deterministic fractal networks are self-similar,  the reciprocal is false: there  exists self-similar networks that do not show the fractal property, such as the so-called  $(u,v)$-flower networks with $u = 1$ that will be defined more precisely below. In the latter example, the diameter scales as a logarithm of the number of nodes: $R \sim \log N$, which is referred to as the small-world property. Such a network can be formally seen as a fractal network with infinite fractal dimension, and is sometimes called transfinite fractal or transfractal \cite{Rozenfeld:2007NJP}.   

Beyond fractality, self-similar networks may also exhibit other prominent properties that seem to be common to real networks, especially biological and social networks, such as scale-free features or modular structure \cite{Ravasz:2002Science}. The former implies that the distribution of the degree of the nodes of the network follows a power law, while the latter means that the network can be divided into groups (modules), within which nodes are more tightly connected with each other than with outside nodes \cite{Girvan:2002PNAS,Ravasz:2003PRE}.
In the last part of this article  we focus on a class of hierarchical, non-decimable, recursive networks able to capture simultaneously scale-free behavior and modular structure, yet preserving (weak) self-similarity \cite{Girvan:2002PNAS,Ravasz:2002Science}. 
As we will see,  the calculation of first-passage quantities on such networks requires an alternative approach that will be presented in the last section of this article. 


This paper is organized as follows. In the first Section, we introduce general definitions and present in detail the method of calculation of splitting probabilities and MFPTs on the example of the Sierpinski gasket. Then,   we extend this method  to other deterministic self-similar networks on the example of the Song-Havling-Makse networks (Section II), while further examples (T graph and $(u,v)$ flowers) are given in Appendix. We stress that this approach allows one to calculate explicitly the splitting probabilities and MFPTs for {\it any} starting site of the lattice (but for specific targets). In addition, we will show that it leads to simple expressions of {\it selective} averages over starting sites (that is averages over specific class of starting sites, to be defined below). In the particular case  where the average is performed over all starting sites, we recover expressions of global MFPTs  recently obtained. Finally, in Section III, we present an  alternative method of calculation in the case of hierarchical networks.

\begin{figure}
\begin{center}
\includegraphics[width=5cm]{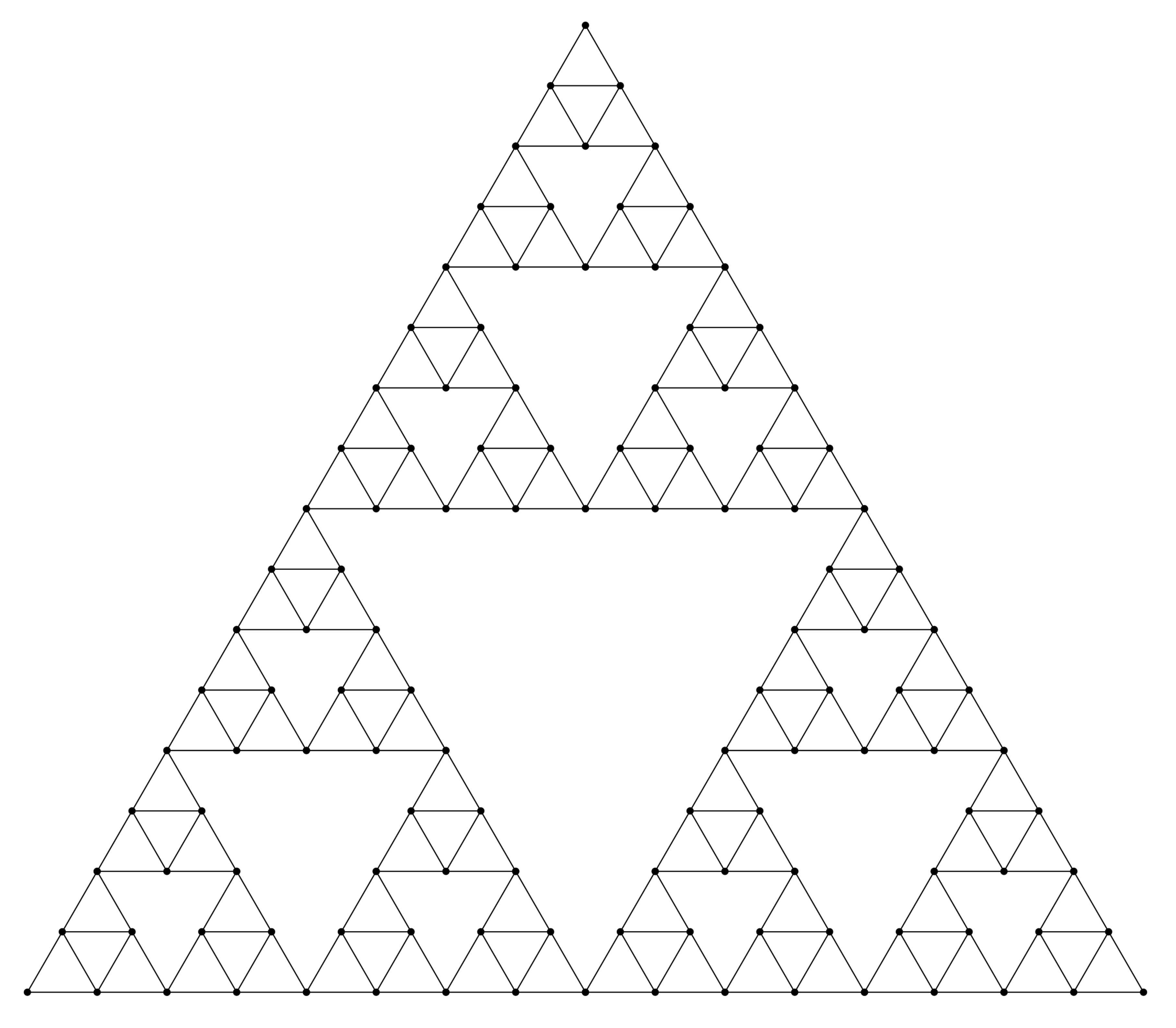}
\includegraphics[width=6.5cm]{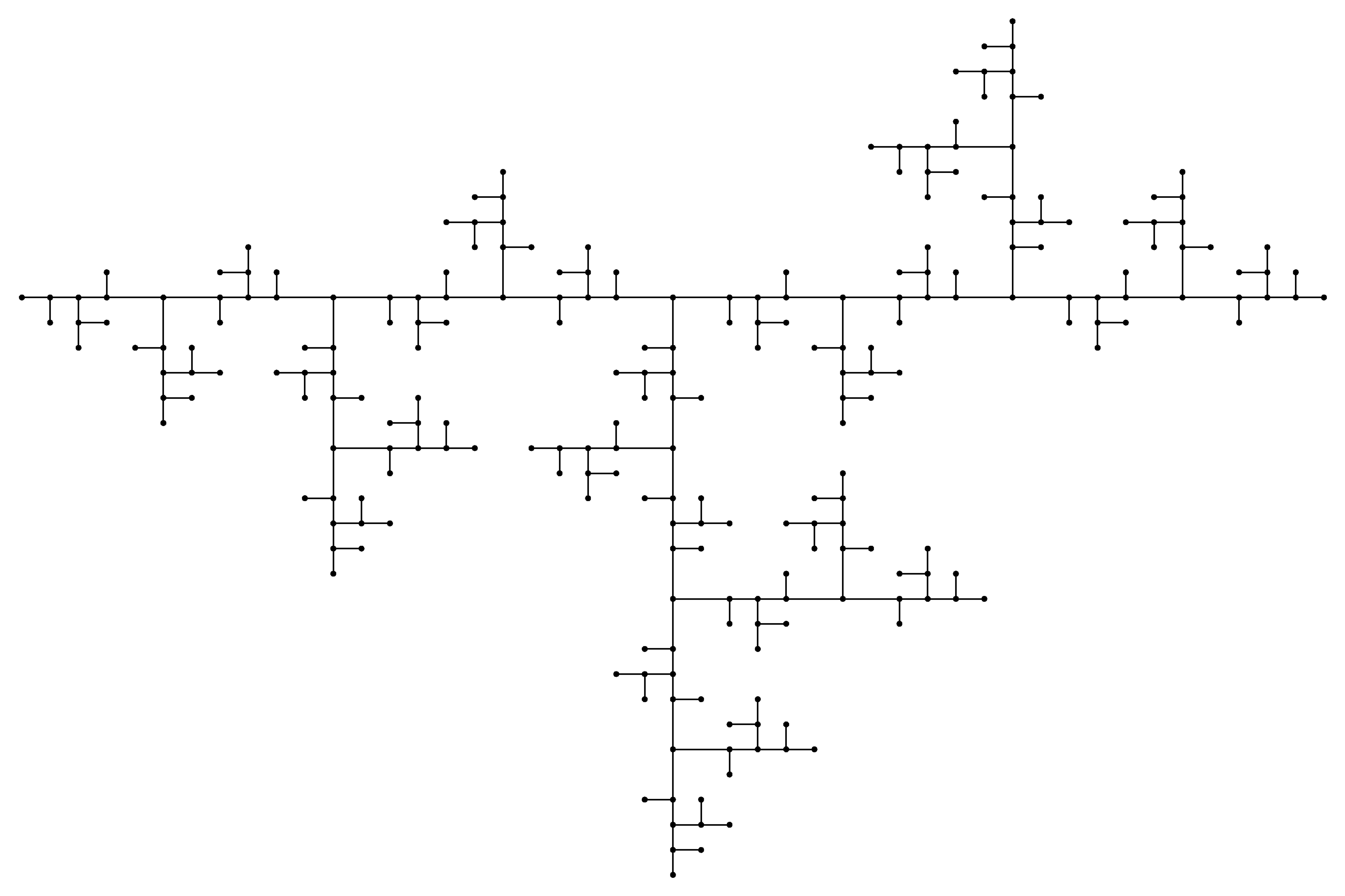}
\caption{\small Examples of self similar networks : 4th generation Sierpinski gasket and 5th generation T-graph. }
\label{fig:ex_fractals}
\end{center}
\end{figure}

\section{The general method illustrated on the example of the Sierpinski gasket}\label{general}

\subsection{Definitions}

 In this section, we first introduce the class of networks to which our method of calculation can be applied, and give basic definitions that will be used throughout the paper.  Note that here we do not aim at giving mathematically formal definitions, and will rather largely rely on explicit examples.

\begin{figure}[h]
\begin{center}
\includegraphics[width=8cm]{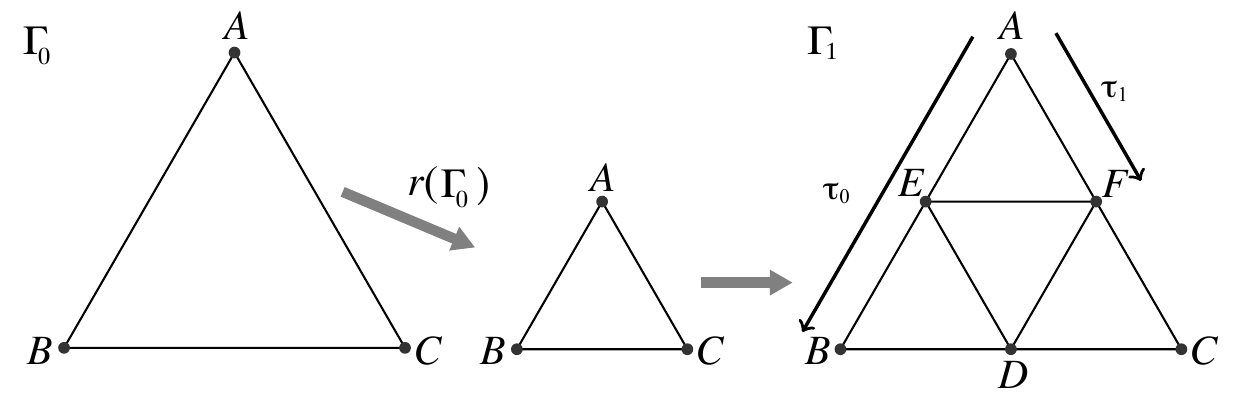}
\caption{\small Sierpinski gasket : renormalization scheme and associated crossing times $\tau_k$.} \label{fig:sierp_tauk_renorm}
\end{center}
\end{figure}

\paragraph*{Hierarchical networks.}  As stated in introduction, we consider hierarchical networks which can be defined \cite{D.Ben-Avraham:2000} by considering the action of   a rescaling transformation $r(\Gamma)$  that maps all points $i$ of a network $\Gamma$  onto new points $i' = r( i)$. A self-similar network is then constructed recursively from an elementary motif (initiator) $\Gamma_0$ by writing $\Gamma_n=\cup_{i=1..p}r(\Gamma_{n-1} ) $. For example, the Sierpinski gasket of generation $n$ is built by joining $p=3$ copies, called subunits, of Sierpinski gaskets of generation $n-1$ (see Figs. \ref{fig:ex_fractals}  and \ref{fig:sierp_tauk_renorm}), where the initiator is the elementary triangular network with three nodes.

\begin{figure}[h]
\begin{center}
\includegraphics[width=5cm]{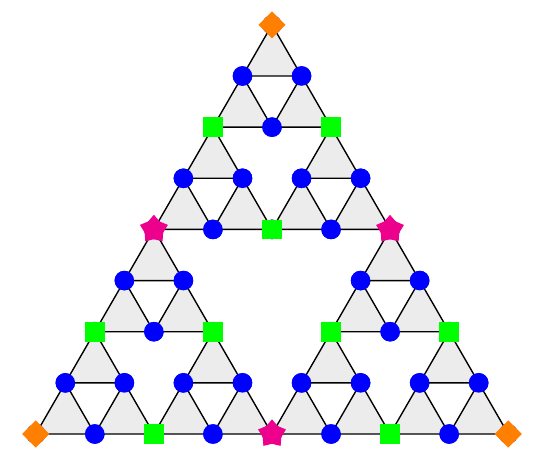}
\caption{\small Color online. Third-generation Sierpinski gasket : levels. Blue circles : level 3 ; green squares: levels 2,3 ; magenta stars: levels 1,2,3 ; orange diamonds: levels 0,1,2,3.} 
\label{fig:sierp_levels}
\end{center}
\end{figure}

\paragraph*{Levels.}  A given site $i_0$ of a hierarchical network $\Gamma_n$ is said to belong  to the level $\xi$, if there exists an integer $\xi \le n$ such that $i_0\in \Gamma_\xi$.  We will denote by ${\cal L}_k$ the set of nodes of level $k$. Note that the networks are constructed such that   a site that belongs to the level $\xi$ also belongs to levels $\xi+1, \;\xi+2,\;...,\;n$ . Figure \ref{fig:sierp_levels} illustrates this definition for the 3rd-generation of a Sierpinski gasket (see legend). Note that, for the $n$th-generation of a Sierpinski gasket, there are $\frac{3^{k+1}+3}{2}$ sites on the level $k$ ($k\in{1,...,n}$).

\paragraph*{Labeling of the sites and subunits.}

\begin{figure}[h]
\begin{center}
\includegraphics[width=8cm]{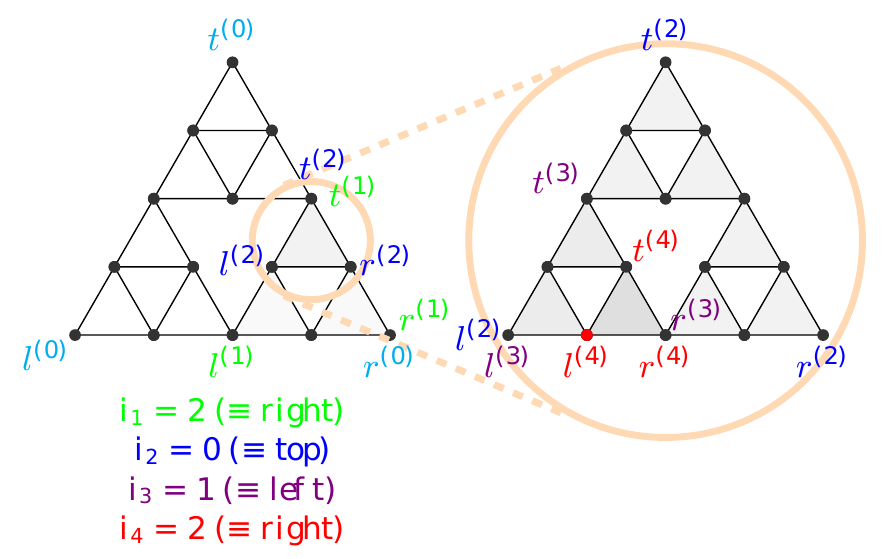}
\caption{\small Color online. Sierpinski gasket : example of labels.} 
\label{fig:sierp_indexes}
\end{center}
\end{figure}

Any subunit of level $k$ in a self-similar network of generation $n$  can be reached recursively by defining a \emph{path}, i.e. a sequence $\{i_0,...,i_k\}$ where $i_j$ labels the position of each of the nodes of the initiator $\Gamma_0$. An example is given in Figure \ref{fig:sierp_indexes} in the case of the Sierpinski gasket, by assigning the value  $i_k=0$ to a top subunit, $i_k=1$ to a left one and 
$i_k=2$ to a right one. With these rules, the path $\{i_1,i_2,i_3,i_4\} = \{2,0,1,2\}$ allows to locate the last-level sub-unit $(t^{(4)},l^{(4)},r^{(4)})$ within the 4th level.

\paragraph*{Transport process.} We consider throughout this paper a nearest neighbor  Markovian random walker characterized by generic transition probabilities  $w({\bf r'}|{\bf r})$ from site $\bf r$ to $\bf r'$. Unless specified we will consider isotropic random walks such that $w({\bf r'}|{\bf r})=1/\kappa ({\bf r})$ where $\kappa({\bf r})$ is the connectivity of node ${\bf r}$.

\subsection{Splitting probabilities}

In this section we wish to calculate the splitting probability $P_{{\bf r}_T|{\bf r}_{A_i}}(\bf r)$, defined as the probability for a Markovian random walker starting at $\bf r$ to reach the target ${\bf r}_T$ in the presence of other absorbing sites $\{{\bf r}_{A_i}\}_i$, in other words the probability to reach the site ${\bf r}_T$ before all the other absorbing sites. For the sake of readability, we will make use of the following notation: 
\be\nonumber
P_{{\bf r}_T|{\bf r}_{A_i}}({\bf r}) = P(\bf r).
\fin

We will present the  method on the example of the Sierpinski gasket and show later on how it can be generalized to any self-similar network. Let us consider a general subunit $\Lambda_{k-1}$ at a given level $k-1$, which is depicted in the right hand side of Fig. \ref{fig:sierp_indexes}. We  assume that the splitting probability $P(\bf r)$  is known for all starting sites of level $k-1$ in $ \Lambda_{k-1}$ (in the Sierpinski gasket there are only 3 such sites which correspond to the summits $A,B,C$ of the main triangle defining $ \Lambda_{k-1}$), and that the absorbing sites ${\bf r}_T$ and $\{{\bf r}_{A_i}\}_i$ are located outside   $\Lambda_{k-1}$. Here the subunit $\Lambda_{k-1}$ is the union of $p=3$ copies of subunits $\Lambda_{k}$.  The splitting probabilities for sites $\bf r$ of level $k$ in $\Lambda_{k-1}$ (namely the sites $A,B,C,D,E,F$ in Fig \ref{fig:sierp_tauk_renorm}) satisfy the following backward equation \cite{Redner:2008}:
\be\label{eq:em_splitting}
0 = \sum_{{\bf r'} \in {\cal L}_{k}\cap \Lambda_{k-1}} \pi({\bf r'}|{\bf r}) P({\bf r'})-P({\bf r}).
\fin
Here $ \pi({\bf r'}|{\bf r})$ is the splitting probability that the random walker starting from the node $\bf r$ of level $k$ in $\Lambda_{k-1}$ reaches first the site $\bf r'$ among all sites of level $k$ in $\Lambda_{k-1}$. The sites of level $k$ in $\Lambda_{k-1}$ actually form a graph $\Gamma_1$ of generation 1 up to a rescaling factor (see Fig \ref{fig:sierp_tauk_renorm}).  In the case of an isotropic walk,   it is clearly seen on  the example of the  Sierpinski gasket (for which all nodes of the initiator $\Gamma_0$ are equivalent) that $ \pi({\bf r'}|{\bf r})= w({\bf r'}|{\bf r})$, where $w({\bf r'}|{\bf r})$ is the elementary transition probability on $\Gamma_1$. More explicitly, on Fig \ref{fig:sierp_tauk_renorm}, on has $ \pi({B}|{ E})= \pi({D}|{ E})= \pi({F}|{ E})= \pi({A}|{ E})=1/4$. In the remainder of the article, we consider isotropic random walks on networks, for which this property holds. Nevertheless, the calculation method that we present can be in principle extended to directed or non-uniformly weighted networks (such that   $ \pi({\bf r'}|{\bf r}) \neq \pi({\bf r'}|{\bf r})$), as long as the scale-invariance hypothesis is fulfilled. 

In this example, let us  assume that $P(A)$, $P(B)$ and $P(C)$ are known; the expressions of  $P(D)$, $P(E)$ and $P(F)$ of the splitting probabilities starting from the points $D,E,F$ of level $k$ in  $\Lambda_{k-1}$ can then be obtained readily by making use of Eq. (\ref{eq:em_splitting}).
 One has
\begin{align}\label{eq:sierp_raw_single_eq}
   P(E) &= \frac{1}{4} \left[ P(A)+P(F)+P(D)+P(B) \right], \\
    P(F) &= \frac{1}{4} \left[ P(A)+P(E)+P(D)+P(C) \right], \\
  P(D) &= \frac{1}{4} \left[ P(B)+P(E)+P(F)+P(C) \right], 
\end{align}
which can be rewritten as a linear system
\begin{equation}\label{eq:sierp_raw_syst}
    \begin{pmatrix} P(D)\\P(E)\\P(F) \end{pmatrix}  =  \begin{pmatrix} 1/5&2/5&2/5\\2/5&2/5&1/5\\2/5&1/5&2/5 \end{pmatrix} \,  \begin{pmatrix} P(A)\\P(B)\\P(C) \end{pmatrix} .
\end{equation}

We now proceed iteratively and consider for example the upper subunit  $\Lambda_k$ of $\Lambda_{k-1}$ that contains the nodes  $A,E,F$ of level $k$.  This choice is taken into account by assigning to  a variable $i_k$ the value $0 \equiv top$. It is of course possible to apply the same operation to any of  the two other subunits : the lower-left one ($i_k = 1 \equiv left$) or the lower-right one ($i_k = 2 \equiv right$).
Let us then rename the nodes of level $k-1$ in  $\Lambda_{k-1}$ and the nodes of level $k$ in  $\Lambda_k$ as follows:
\begin{eqnarray*}
 && t^{(k-1)} = A, ~l^{(k-1)} = B, ~r^{(k-1)} = C, \\
 && t^{(k)} = A, ~l^{(k)} = E, ~r^{(k)} = F .
\end{eqnarray*}
Equation  (\ref{eq:sierp_raw_syst}) can then be rewritten as: 
\begin{equation}\label{eq:sierp_rec_top}
  \mathcal{P}^{(k)}\equiv \left( \begin{array}{c} P(t)\\P(l)\\P(r) \end{array} \right)^{(k)}  =  \begin{pmatrix} 1&0&0\\2/5&2/5&1/5\\2/5&1/5&2/5 \end{pmatrix} \, \left( \begin{array}{c} P(t)\\P(l)\\P(r) \end{array} \right)^{(k-1)}.
\end{equation}

Similarly, one can define three matrices $\mathcal{M}_{i_k}$ such that  equation (\ref{eq:sierp_rec_top}) reads for each value  of $i_k$ corresponding to either a top, left or right subunit: 
\be\label{eq:sierp_sp_recursive}
\mathcal{P}^{(k)} =  \mathcal{M}_{i_k} \,\mathcal{P}^{(k-1)}
\fin
with
\[
{\cal M}_0 = \begin{pmatrix} 1&0&0\\2/5&2/5&1/5\\2/5&1/5&2/5 \end{pmatrix}, \ \ \
{\cal M}_1 = \begin{pmatrix} 2/5&2/5&1/5\\0&1&0\\1/5&2/5&2/5 \end{pmatrix} 
\]
\be \label{eq:sierp_matrices}
{\rm and} \ \ {\cal M}_2 = \begin{pmatrix} 2/5&1/5&2/5\\1/5&2/5&2/5\\0&0&1 \end{pmatrix}.
\fin

 As shown above, any subunit (of level $k$) in the network can be reached recursively by defining a \emph{path}, i.e. a sequence $\{i_1,...,i_k\}$ with $i_l \in \{0,1,2\} $ for $ 1 \le l \le k$. Iterating equation (\ref{eq:sierp_sp_recursive}) then yields straightforwardly: 
\be\label{eq:sierp_sp_nonrecursive}
\mathcal{P}^{(k)} =  \mathcal{M}_{i_k} \mathcal{M}_{i_{k-1}} \cdots \mathcal{M}_{i_{k_0+1}} \,\mathcal{P}^{(k_0)}.
\fin 
This shows that as soon as  $\mathcal{P}^{(k_0)}$ is known for a given level $k_0$, the splitting probability starting from {\it any} point of level $k\ge k_0$ can be obtained exactly and only requires to compute a product of $k-k_0$ $3\times3$ matrices. In particular if the targets are chosen among the 3 sites of level 0 then  $\mathcal{P}^{(k_0)}$ is calculated  trivially and the splitting probability for any starting point of the network is readily obtained. 


Let us consider an explicit example illustrated in  Fig. \ref{fig:sierp_indexes} that represents a  Sierpinski gasket of generation 4. We aim at  calculating the splitting probability to reach $t^{(0)}$ before $r^{(0)}$  starting from $l^{(4)}$: 
\[
 P_{t^{(0)}|r^{(0)}} (l^{(4)}) \equiv P(l^{(4)}). 
\]
$ P(l^{(4)})$ is the 2nd coordinate of the vector $\mathcal{P}^{(4)}$, associated to the subunit of level 4 (with vertices in red). Since the three matrices ${\cal M}_0$, ${\cal M}_1$ and ${\cal M}_2$  are known, all we need to determine is the path of that subunit and the value of $\mathcal{P}^{(0)}$. As before, one has:
\begin{eqnarray*}
 {i_1,i_2,i_3,i_4} & = &{2,0,1,2}  
\end{eqnarray*}
and clearly
\begin{eqnarray*}
  \mathcal{P}^{(0)} &=& \begin{pmatrix} 1\\1/2\\0 \end{pmatrix}.
\end{eqnarray*}

Thus, applying equation (\ref{eq:sierp_sp_nonrecursive}), we get $\mathcal{P}^{(4)} = \begin{pmatrix} 234/625\\229/625\\89/250 \end{pmatrix}$, so that finally:
\be
   P(l^{(4)}) = \frac{229}{625} 
\fin

\subsection{MFPT}\label{sec:sierp_MFPT}

In this section we consider the MFPT  of a random walker to a target site ${\bf r}_T$ starting from a site $\bf r$, that we denote $T({\bf r})$. Following the main steps of derivation of the splitting probability, we first write down a backward equation for the MFPT starting from a given site $\bf r$ of level $k$ in a subunit $\Lambda_{k-1}$ of level $k-1$ (see Fig. \ref{fig:sierp_tauk_renorm}): 
\be\label{eq:em_mfpt}
-\tau_k=\Delta T({\bf r}) = \sum_{{\bf r'} \in {\cal L}_{k}\cap \Lambda_{k-1}} \pi({\bf r'}|{\bf r}) T({\bf r'})-T({\bf r}).
\fin
As discussed before, the splitting probability $ \pi({\bf r'}|{\bf r})$ is readily given by the transition probability  $w({\bf r'}|{\bf r})$ on the corresponding graph $\Gamma_1$ of generation 1 formed by the sites of level $k$ in  $\Lambda_{k-1}$ (sites $A,B,C,D,E,F$ in Fig \ref{fig:sierp_tauk_renorm}).  In addition we introduced the quantity $\tau_{k}$ defined as the time it takes a random walker to exit a subunit of level $k$. On the example of the right hand side of Fig \ref{fig:sierp_tauk_renorm}, $\tau_{k}$ is the mean time to reach either $B,D,F,A$ starting from $E$. Note that by symmetry it can also be defined as the mean time to reach either $E$ or $F$ starting from $A$. In this example, all the exit nodes of a given subunit play a symmetric role,  and by construction, all subunits of a given level are the same. Therefore $\tau_{k}$ does not depend on ${\bf r}$ 
in equation (\ref{eq:em_mfpt}). This property results here from the symmetry of the initiator $\Gamma_0$, in which all nodes are equivalent, and from the symmetry of the transition probabilities.  A stated previously, we will consider in this paper only networks having this property.

More explicitly, we rely on the example of Fig. \ref{fig:sierp_tauk_renorm} and assume that the MFPT starting from the sites $A,B,C$ of level $k-1$ is known. Using Eq. (\ref{eq:em_mfpt}), one can write :
\begin{eqnarray*}
  T(E) & = &\frac{1}{4} \Big[ (T(A) + \tau_k) + (T(F) + \tau_k) \\
  && + (T(D) + \tau_k) + (T(B) + \tau_k) \Big],
\end{eqnarray*}
and the the 2 similar relations at nodes $F$ and $D$.  
Following the derivation of splitting probabilities above, the MFPT starting from any site of level $k$ can be expressed linearly in terms of the MFPT starting from sites of level $k-1$. For example, focusing on the top subunit $AEF$ (corresponding to  the choice $i_k = 0  $), one obtains:
\begin{eqnarray}
 \mathcal{T}^{(k)} & = & \tau_k \, \begin{pmatrix} 0\\2\\2 \end{pmatrix} + \begin{pmatrix} 1&0&0\\2/5&2/5&1/5\\2/5&1/5&2/5 \end{pmatrix} \, \mathcal{T}^{(k-1)}  \nonumber \\
  & = & \tau_k \, \mathcal{V}_0  + \mathcal{M}_0 \, \mathcal{T}^{(k-1)},
\end{eqnarray}
where $\mathcal{T}^{(k)} \equiv  \begin{pmatrix} T(t)\\T(l)\\T(r) \end{pmatrix}^{(k)}$ denotes the vector  of MFPTs  starting from the 3 vertices of level $k$ of a given subunit of level $k$ located by its path $\{i_1,...,i_k\}$. Similar equations can be obtained for   the left ($i_k = 1$) or the right ($i_k = 2$) subunit, and yield  the following general recursive relation : 
\be\label{eq:sierp_mfpt_recursive}
\mathcal{T}^{(k)} = \tau_k \, \mathcal{V}_{i_k} +  \mathcal{M}_{i_k} \,\mathcal{T}^{(k-1)},
\fin
where the $\mathcal{M}_{i_k}$ matrices are given by Eq. (\ref{eq:sierp_matrices}) and
\be\label{eq:sierp_vectors}
\mathcal{V}_0 = \begin{pmatrix} 0\\2\\2 \end{pmatrix} ~~~\mathcal{V}_1 = \begin{pmatrix} 2\\0\\2 \end{pmatrix}~~~\mathcal{V}_2 = \begin{pmatrix} 2\\2\\0 \end{pmatrix}.
\fin

Assuming that  $\mathcal{T}^{(k_0)}$ is known for some level $k_0$, the MFPT for a given level $k>k_0$ can then be written as:
\be \label{eq:sierp_mfpt_nonrecursive}
\mathcal{T}^{(k)} =\tau_k \, \mathcal{V}_{i_k} + \sum_{l=k_0}^{k-1} \tau_{l} \left( {\cal M}_{i_k}{\cal M}_{i_{k-1}} \cdots {\cal M}_{i_{l+1}} \right) {\cal V}_{i_l} ,
\fin
where here by definition $ \tau_{i_{k_0}} {\cal V}_{i_{k_0}}\equiv \mathcal{T}^{(k_0)}$. We give below the example where the target is located at the apex, for which the determination of $\mathcal{T}^{(0)}$ is straightforward. As we show in the next paragraph, $\tau_k$ can be calculated explicitly. In the case where the target is located at the apex, Eq. (\ref{eq:sierp_mfpt_nonrecursive}) therefore provides an explicit and exact expression of the MFPT starting from any starting site of the network. Examples will be given below.   

\paragraph*{Determination of the exit time $\tau_k$.}


The notations we refer to are those of figure \ref{fig:sierp_tauk_renorm}. It is easily seen that  $\tau_k$ and $\tau_{k-1}$ are related by the following system: 
\[
 \left\{
\begin{array}{l}
\tau_{k-1} = \tau_k + \frac{1}{2} (\mathfrak{T}(E)+\mathfrak{T}(F))  \\
\mathfrak{T}(E) = \frac{1}{4} \left[ (\tau_k + \tau_{k-1} ) + (\tau_k + \mathfrak{T}(F)) + (\tau_k + \mathfrak{T}(D)) + \tau_k \right] \\
\mathfrak{T}(D) = \frac{1}{4} \left[ (\tau_k + \mathfrak{T}(E)) +(\tau_k + \mathfrak{T}(E)) \right] + \frac{1}{2} \tau_k  \\
\mathfrak{T}(E) = \mathfrak{T}(F), 
\end{array}
\right.
\]
where $\mathfrak{T}(E)$, $\mathfrak{T}(F)$  and $\mathfrak{T}(D)$ denote the mean time to reach \emph{any} of the two $B$ or $C$ sites starting from $E$, $F$ and $D$. The solution of this system is $
\tau_{k-1} = 5\, \tau_k$
which leads to $ \tau_k = 5^{n-k} \tau_n$. At level $n$ the subunit is a simple  triangular graph so that obviously $\tau_n = 1$. Finally, one has:
\be
\tau_k = 5^{n-k}.
\fin

\paragraph*{Example 1.} Here, we refer again to Fig. \ref{fig:sierp_indexes} and we wish to calculate $T(l^{(4)})$ when the target is set at the apex $t^{(0)}$. First, it is necessary to determine  $\mathcal{T}^{(0)}$: 
\[
T(l^{(0)}) =  \tau_0 + \frac{1}{2}\,T(r^{(0)})  \ \ {\rm and} \ \ T(l^{(0)}) = T(r^{(0)}),  
\]
therefore
\be
\mathcal{T}^{(0)} = 2\times5^n\begin{pmatrix} 0\\1\\1 \end{pmatrix}  =   \tau_0 \mathcal{V}_0.
\fin
The path and the value of $\tau_k$ are already known. Finally one obtains:
\be
  \mathcal{T}^{(4)} = 1176
\fin

\paragraph*{Example 2 : a class of sources.} In the present example, the aim is to study the MFPT dependence with respect to the target-source distance. The target is left at $t^{(0)}$, and the class of sources is the sites located  on the adjacent edge [$t^{(0)},l^{(0)}$], at the distances $r_p=2^p$, $0 \le p \le n$.
Those sources are the lower-left vertices of the triangles corresponding to the paths $\{0\}$ ($r=2^n$), $\{0,0\}$ ($r=2^{n-1}$), $\{0,0,0\}$ ($r=2^{n-2}$), etc. Thus, the problem is solved by using formula (\ref{eq:sierp_mfpt_nonrecursive}) for $k=n-p$ and $\mathcal{M}_{i_l} = \mathcal{M}_{0}$ for all $l$. We find finally:
\be\label{exactsirp}
T(r=2^p) = 5^p \,(3^{n-p+1}-1).
\fin

The latter result can be compared to the general asymptotic  expression of the MFPT  that has recently been derived in \cite{Benichou:2008PRL} in the large system size limit. In the case of a target located at the apex of the Sierpinsky gasket it reads for $N$ large : 
\be
 \,T_a \sim 2N r^{d_w-d_f},
\fin
where  $r$  is the  source-target distance, $d_f$ the fractal dimension of the network and $d_w$ the walk dimension, characterizing the power-law behaviour  of the mean-square displacement with respect to time : $\langle \Delta r^2 \rangle \sim t^{2/d_w}$. In the present case, $2N=3^{n+1}+3$, $d_f = \ln(3)/\ln(2)$ and $d_w=\ln(5)/\ln(2)$. Thus:
\be
T_a \sim 5^p\, 3^{n-p+1}.
\fin
As expected this asymptotic regime is recovered by taking the large volume limit $n\to\infty$ in the exact expression (\ref{exactsirp}).

\subsection{Averages}\label{sec:mfpt_avg}

In this section, we aim at calculating the MFPT to  a target site averaged over different  classes of starting points. The average can cover either a class of starting points (all the sites of a given level, all sites of a given connectivity...) or all the sites of network. In this latter case the averaged MFPT is often called the  Global MFPT \cite{Kozak:2002PRE,Haynes:2008PRE}. We will show in this section that averages of the MFPT over all sites of a given level take simple explicit forms.

In the case of the Sierpinski gasket, we start from  Eq. (\ref{eq:sierp_mfpt_nonrecursive}) that gives an explicit expression of the MFPT starting from any of the three points of level $k$ in a given subunit of level $k$. Since each subunit of level $k$ is in one to one correspondence with a path  $\{i_1,\cdots,i_k\}$, one has to calculate 
 $\sum_{\{i_1,\cdots,i_k\}} \mathcal{T}^{(k)}$. Here we assume that the target site is at level 0 and that $\mathcal{T}^{(0)}$ is known. Making use of the following identity
\[
\sum_{\{i_1,\cdots,i_k\}} \sum_{l=0}^k = \sum_{l=0}^k \sum_{\{i_1,\cdots,i_k\}},
\]
we are back to calculate the expression 
\[
 \sum_{\{i_1,\cdots,i_k\}}  \left( {\cal M}_{i_k}{\cal M}_{i_{k-1}} \cdots {\cal M}_{i_{l+1}} \right) {\cal V}_{i_l}.
\]
 The variable  $i_k$ is chosen out of a set of $p$ values (in the case of the Sierpinski gasket, $p=3$), and one has:
\begin{eqnarray}
&& \sum_{\{i_1,\cdots,i_k\}}  \left( {\cal M}_{i_k}{\cal M}_{i_{k-1}} \cdots {\cal M}_{i_{l+1}} \right) {\cal V}_{i_{l}} =p^{l-1}   \nonumber \\
&& \times \left( {\cal M}_0 + {\cal M}_1 + ... + {\cal M}_p \right)^{k-l} \left( {\cal V}_0 + {\cal V}_1 + ... + {\cal V}_p \right). \nonumber \\
&&
\end{eqnarray}
We then define ${\cal M}_{tot} = {\cal M}_0 + ...+ {\cal M}_p$ and ${\cal V}_{tot} = {\cal V}_0 + ...+ {\cal V}_p$, and obtain the following general formula:
\be\label{eq:mfpt_avg_general}
\sum_{paths} \mathcal{T}^{(k)} = (\mathcal{M}_{tot})^k \, \mathcal{T}^{(0)} + \sum_{l=1}^k \tau_l \, p^{l-1} (\mathcal{M}_{tot})^{k-l} \, \mathcal{V}_{tot}.
\fin
We give below explicit examples. 

\paragraph*{Case of a single target at the apex. }  Let us consider a given level $k$, and denote by   $S^{(k)}$  the MFPT to a target site set at the apex summed over all the starting sites of level $k$. 
Using expressions (\ref{eq:sierp_matrices}) and (\ref{eq:sierp_vectors}), and the fact that $p=3$ for the Sierpinski gasket, the diagonalization  of  ${\cal M}_{tot}$ leads to:
\begin{eqnarray}\label{eq:sierp_mfpt_avg_terms}
p^{l-1} \left( {\cal M}_{tot} \right)^{k-l} {\cal V}_{tot} 
& = & \frac{3^{k-1}}{5^{k-l}} \begin{pmatrix} 3&1&1\\1&3&1\\1&1&3 \end{pmatrix}^{k-l} \begin{pmatrix} 4\\4\\4 \end{pmatrix} \nonumber \\
& =&  4\times 3^{k-1}\begin{pmatrix} 1\\1\\1 \end{pmatrix}.
\end{eqnarray}
Then, using $ \mathcal{T}^{(0)}= \tau_0 {\cal V}_0$, Eq. (\ref{eq:mfpt_avg_general}) yields:
\begin{eqnarray}\label{sumpaths}
\sum_{paths} \mathcal{T}^{(k)} &=& 4\times 6^{k-1}\,5^{n-k} \begin{pmatrix} -2\\1\\1 \end{pmatrix} \nonumber \\
&& + \, 5^{n+1}\,3^{k-1} \left(1-5^{-k-1}\right) \begin{pmatrix} 1\\1\\1 \end{pmatrix}.
\end{eqnarray}

 The first coordinate of the latter expression corresponds to the sum of the MFPTs starting from the top vertex of each of the subunits of level $k$. Its 2nd and 3rd coordinates are respectively the sum for the left vertices and for the right vertices. An examination of the Sierpinski gasket shows that the sum of  the three coordinates of the  vector defined by Eq. (\ref{sumpaths}) is equal to $2\,S^{(k)} - T(l^{(0)}) - T(r^{(0)})$. 
 Knowing that the number of sites of level $k$ is $\frac{3^{k+1}+1}{2}$, we finally obtain the MFPT ${\overline{\aver{T}}}^{(k)}$ averaged over all starting sites of level $k$: 
 \begin{eqnarray}\label{eq:tmoysierp}
{\overline{\aver{T}}}^{(k)} & = &  \frac{2 \,S^{(k)}}{3^{k+1}+1}  \nonumber\\
& = & \frac{5^{n+1}\,3^k + 4\times5^n - 5^{n-k}\,3^k}{3^{k+1}+1} . 
\end{eqnarray}
   For the particular case $k=n$ this quantity is the GMFPT and is in agreement  with the results of \cite{Kozak:2002PRE}.

\paragraph*{The 3 vertices of level 0 are targets.}  We now assume  that the 3 vertices of  level 0 are targets and wish to calculate the MFPT averaged over all starting points of level $k$. The only quantity that we need to modify in the calculation of the previous paragraph is $\mathcal{T}^{(0)}$. In the case of 3 targets one has straightforwardly
\be
\mathcal{T}^{(0)} = \begin{pmatrix} 0\\0\\0 \end{pmatrix}.
\fin
The other terms of equation (\ref{eq:mfpt_avg_general}), that are given by equation (\ref{eq:sierp_mfpt_avg_terms}), are unchanged. We thus have: 
\be
\sum_{paths} \mathcal{T}^{(k)} = 5^n\,3^{k-1} \left(1-5^{-k}\right) \begin{pmatrix} 1\\1\\1 \end{pmatrix}.
\fin
We finally obtain the desired quantity:
\be
{\overline{\aver{T}}}^{(k)} = \frac{5^n\,3^{k-1} (1-5^{-k})}{3^{k+1}+1}.
\fin

\subsection{Conclusion}
In this section we have derived on the example of the Sierpinski gasket exact expressions of the splitting probabilities  and MFPTs. Our method yields simple expressions in the case where the target(s) belong(s) to the level 0 of the network and applies to any starting site. This method can be readily generalized to other self-similar networks for which a similar addressing of the subunits of each level can be defined. The only task is then to calculate the matrices $ {\cal M}_{i}$, the vectors $ {\cal V}_{i}$ and the exit time $\tau_k$. The example of the T--graph is given in appendix, and further examples that require slight  adaptations of the method are detailed below.

\section{The Song-Havlin-Makse networks}

\begin{figure}
\begin{center}
\includegraphics[height=5cm]{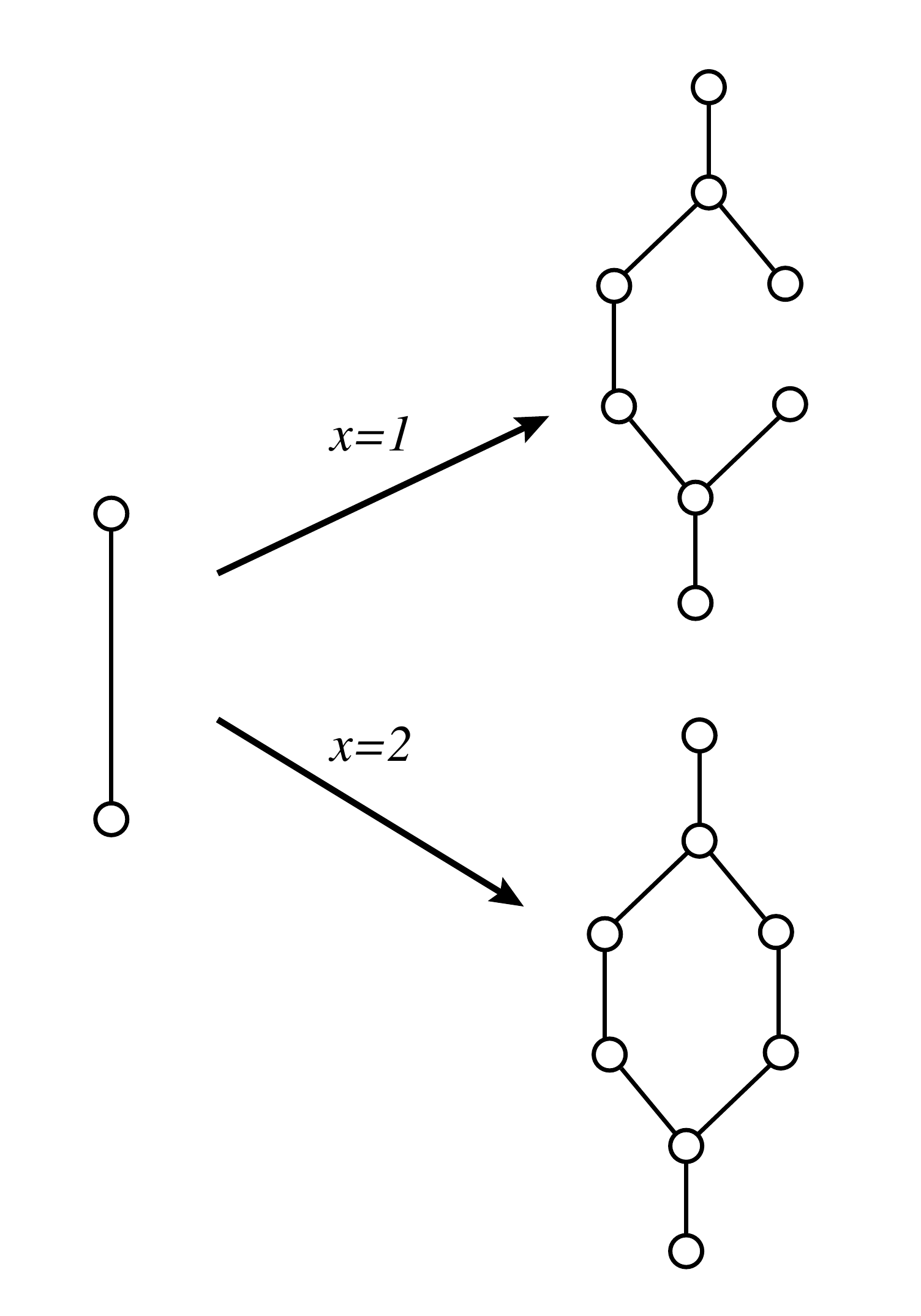}
\caption{\small Song-Havlin-Makse network renormalization scheme: starting from a link between two nodes, the next generation is obtained by attaching $m$ new 
nodes (sons) to each of them, then  by deleting the original link and by creating $x$ new links connecting the newly created sons. Example of $m=3$ and $x=1$ or $x=2$.   }
\label{fig:renorm_song}
\end{center}
\end{figure}

We now consider  the example of the Song-Havlin-Makse networks, which have been introduced in  \cite{Song06NatPhys,Gallos07PNAS} in order to build self-similar networks that mix fractal and non-fractal growing schemes. Here,  we  focus on the deterministic  version of these networks, and show that the method developed in the previous section applies upon minor modifications.

The  building scheme of these networks  starts from a single link between two nodes. The next generation is obtained by attaching $m$ new 
nodes (sons) to each of them, then   by deleting the original link and by creating $x$ new links connecting the newly created sons, as shown on figure \ref{fig:renorm_song}. 
These networks are self-similar and a labeling of each subunit of level $k$ can be defined. As before we will derive the corresponding  matrices $\mathcal{M}_{i_k}$, the vectors $\mathcal{V}_{i_k}$ and the exit time $\tau_k$. It is then possible to calculate explicitly splitting probabilities and MFPTs for any starting point on the network;  here we will calculate examples of MFPTs averaged over different classes of sources.

\begin{figure}
\begin{center}
\includegraphics[height=5cm]{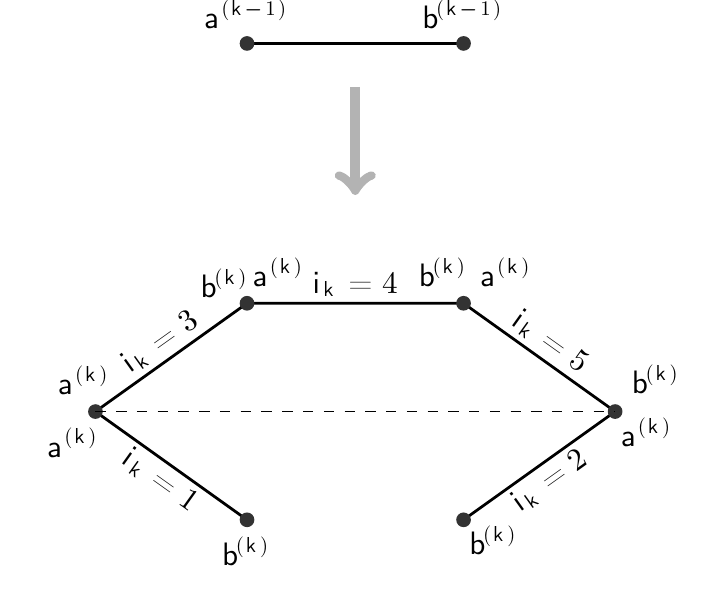}
\caption{\small  Song-Havlin-Makse network : indexes. The indexation of the points does not depend on $m$ nor $x$ (when $x \neq 1$), because all the $x$ connection links play a symmetric role, likewise the $m-x$ free links. Note the switch of positions of $a$ and $b$ in the case $i_k = 2$.} 
\label{fig:song_indexes}
\end{center}
\end{figure}

\paragraph*{Labeling of subunits and sites.} Figure \ref{fig:song_indexes} shows a possible labeling based on 2-dimensional vectors. The scheme does not depend on $m$ and $x$ (provided that $x \ge 1$). This is due to the symmetric role of the $x$ connecting links (respectively of the $m-x$ free links). Although the results do depend on $m$ and $x$, this dependence does not apply to the matrices and vectors. This can be understood as follows. By definition (see Eq.  (\ref{eq:sierp_sp_recursive})), ${\cal M}_k$ determines how the splitting probabilities $P(a^{(k)})$ and $P(b^{(k)})$ at level $k$ are related to the same quantities at level $k-1$. It is easy to see that all trajectories starting  from one of the nodes at level k in one branch will reach sites  $a^{(k-1)}$ or $b^{(k-1)}$ of level $k-1$ before any other site of other branches. As a consequence, ${\cal M}_k$ is independent of the number $x$ of linking branches (or the number $m-x$ of free branches).

\paragraph*{Matrices and vectors.} One finds

\[
{\cal M}_1 = \begin{pmatrix} 1&0\\1&0 \end{pmatrix}, ~~ {\cal M}_2 = \begin{pmatrix} 0&1\\0&1 \end{pmatrix}, ~~ {\cal M}_3 = \begin{pmatrix} 1&0\\2/3&1/3 \end{pmatrix},
\]
\be
   {\cal M}_4 = \begin{pmatrix} 2/3&1/3\\1/3&2/3 \end{pmatrix}, ~~{\cal M}_5 = \begin{pmatrix} 1/3&2/3\\0&1 \end{pmatrix},
\fin
\be
{\cal V}_1 = \begin{pmatrix} 0\\1 \end{pmatrix}, {\cal V}_2 = \begin{pmatrix} 0\\1 \end{pmatrix},  {\cal V}_3 = \begin{pmatrix} 0\\2 \end{pmatrix}, {\cal V}_4 = \begin{pmatrix} 2\\2 \end{pmatrix},  {\cal V}_5 = \begin{pmatrix} 2\\0 \end{pmatrix}. 
\fin

\paragraph*{Crossing time.} One finds
\be
\tau_k = \left( 3 + \frac{6m}{x} \right) ^{n-k} .
\fin

\subsection{MFPT averaged over starting sites of level $k$} \label{sec:song_avg_same_level}

The target is set at $a^{(0)}$. Let us introduce the parameter $\nu_{i_k}$, that numbers the repetition of a branch $i_k$ :
\[
  \left\{ \begin{array}{l} \nu_1 = \nu_2 = m-x \\ \nu_3 = \nu_4 = \nu_5 = x \end{array} \right..
\]
Due to the branch repetitions, it is necessary to adapt equation (\ref{eq:mfpt_avg_general}) by using the following notation :
\[
 \mathcal{M}_{tot} = \sum_{q=1}^5 \nu_q \, \mathcal{M}_q ~;~~\mathcal{V}_{tot} = \sum_{q=1}^5 \nu_q \, \mathcal{V}_q ~;~~\mathcal{\nu}_{tot} = \sum_{q=1}^5 \nu_q.
\]
The analog of equation (\ref{eq:mfpt_avg_general}) is then :
\be\label{eq:song_raw_sum_paths}
\sum_{paths} \mathcal{T}^{(k)} = \tau_0 \, \mathcal{M}_{tot}^k \, \mathcal{V}_{i_0}^{} + \sum_{l=1}^k \tau_l \, \nu_{tot}^{l-1} \mathcal{M}_{tot}^{k-l} \, \mathcal{V}_{tot},
\fin
with ${\cal V}_{i_0} = \begin{pmatrix} 0\\1 \end{pmatrix}$. Let us define: 
\[
 \mu = 3 + \frac{6m}{x}, ~~~~ \lambda_1 = x,  ~~~~ \lambda_2 = 2m+x
\]
where $\lambda_1$ and $\lambda_2$ are the two eigenvalues of $\mathcal{M}_{tot}$. Thus: 
 \begin{align}\label{eq:song_sum_calT_vect}
 \sum_{paths}  \mathcal{T}^{(k)}  = &\, \frac{1}{2} \mu^n \left[\lambda_1^k \begin{pmatrix} -1\\1 \end{pmatrix} +\lambda_2^k \begin{pmatrix} 1\\1 \end{pmatrix}\right] + \, \sum_{l=1}^k \mu^{n-l}\lambda_2^{l-1} \nonumber \\
  \times& \left[ \lambda_1^{k-l}  (m-x) \begin{pmatrix} -1\\1 \end{pmatrix} + \lambda_2^{k-l} (m+3x) \begin{pmatrix} 1\\1 \end{pmatrix} \right] \nonumber\\
   = \, \bigg[ \frac{1}{2}& \left(1+\frac{m-x}{\lambda_2}\right) \lambda_1^k \begin{pmatrix} -1\\1 \end{pmatrix}\nonumber \\
   +  \bigg(&\frac{1}{2} + \frac{m+3x}{(\mu-1)\lambda_2}\bigg) \lambda_2^k \begin{pmatrix} 1\\1 \end{pmatrix} \nonumber \\ 
   - \bigg(&\frac{m-x}{2\lambda_2}  \begin{pmatrix} -1\\1 \end{pmatrix} +\frac{m+3x}{(\mu-1)\lambda_2}  \begin{pmatrix} 1\\1 \end{pmatrix} \bigg) \left(\frac{x}{3}\right)^k  \bigg] \mu^n.
 \end{align}

In order to relate the previous formula to ${\overline{\aver{T}}}^{(k)}$, we need to take into account the repetitions of branch $i_k=5$, otherwise the nodes belonging to the levels $\le k-1$ would be counted several times. Let us call $N_k$ the number of sites belonging to level $k$. An examination of the labeling scheme shows that  ${\overline{\aver{T}}}^{(k)}$ is related to the 2nd coordinate of $ \sum_{paths} \mathcal{T}^{(k)}$ by:
\be\label{eq:song_substracting_levels}
 \left(N_k-1\right) {\overline{\aver{T}}}^{(k)} = \sum_{paths} \left. \mathcal{T}^{(k)} \right|_2  - (x-1)\, \sum_{l=0}^{k-1} \sum_{paths} \left. \mathcal{T}^{(l)} \right|_2 ,
\fin
which yields:
\begin{itemize}
 \item for $x\neq3$:
 \begin{align}\label{eq:song_mfpt_avg_1}
   \overline{\aver{T}}^{(k)}   = &\, \frac{\mu^n}{N_k-1}\Bigg[ \frac{1}{2} \left(1+\frac{m-x}{\lambda_2}\right) \nonumber\\ 
   & +  \left(\frac{1}{2} + \frac{m+3x}{(\mu-1)\lambda_2}\right) \frac{\lambda_2^{k+1}-x\lambda_2^k+x-1}{\lambda_2-1} \nonumber \\
 & - \, \bigg(\frac{m-x}{2\lambda_2}  +\frac{m+3x}{(\mu-1)\lambda_2} \bigg)  \, \frac{3x-3-2x\left(x/3\right)^k}{x-3} \Bigg], 
 \end{align}
\item for $x = 3$: in the latter expression the fraction $\frac{3x-3-2x\left(x/3\right)^k}{x-3}$ (in the third term of the bracket) needs to be replaced by $1-2k$.
\end{itemize}
In expression (\ref{eq:song_mfpt_avg_1}) the volume of level $k$ writes :
\be
N_k = 2m \frac{ \lambda_2^k-1}{\lambda_2-1}+2.
\fin

\subsection{MFPT averaged over the starting  sites of  connectivity 1.} 

Let us keep the target at $a^{(0)}$, and calculate the MFPT averaged over the starting sites of connectivity equal to 1. They belong to level $n$; let us call $N_{\kappa=1}$ their number. One has:
\begin{equation}
  N_{\kappa=1} = 2\,(m-x)\,\lambda_2^{n-1}.
\end{equation}
 An examination of the labeling scheme leads to the following formula, that relates the sum   $ \left(N_{\kappa=1}-1\right)\overline{\aver{T}}_{\kappa=1}$  to the coordinates of $ \sum_{paths} \mathcal{T}^{(k)}$, which has been calculated previously:
\begin{align}\label{eq:song_kappa_1}
 \left(N_{\kappa=1}-1\right)&\, {\overline{\aver{T}}}_{\kappa=1} =\, \sum_{paths} \left(\left. \mathcal{T}^{(n)} \right|_2  -\left. \mathcal{T}^{(n)} \right|_1 \right) \nonumber\\ 
  &+\,   \sum_{paths} \left(m\left. \mathcal{T}^{(n-1)} \right|_1 - (2x-m)  \left. \mathcal{T}^{(n-1)} \right|_2 \right).
\end{align}

The MFPT averaged over all sites of connectivity 1 can then be calculated, using expression (\ref{eq:song_sum_calT_vect}):
\begin{align}
  \left(N_{\kappa=1}-1\right) {\overline{\aver{T}}}_{\kappa=1} =&\, \bigg[ 4\lambda_2^{n-1}\left(3m+2x+\frac{3m^2}{x}\right) \nonumber \\
 +&\, \frac{2}{3}\left(\frac{x}{3}\right)^{n-1}\left(3m-7x\right) \bigg]\frac{ \mu^n (m-x)}{\lambda_2(\mu-1)}.
\end{align}

We show in appendix that this method also applies to the case of the $(u,v)$-flower networks introduced in \cite{Rozenfeld:2007PRE} as examples of deterministic scale-free
networks, that are either fractal  or small-world.

\section{A different method for hierarchical graphs}
\subsection{Recursivity and Modularity}
We now consider a different class of graphs, that is hierarchical, non-decimable, self-similar networks, which are built deterministically and recursively in a manner reminiscent of exact fractal lattices. More precisely, the graph of generation $g$ is obtained by properly \emph{linking} together a certain number of copies of generation $g-1$. Differently from networks previously analyzed, where different replicas meet at a single node, here exact renormalization procedures are not applicable. Yet, we can exploit modularity to detect analogous subgraphs whose nodes satisfy intrinsic, mutual relations, and self-similarity, which allows to establish recursion relations. 

\begin{figure*}
 \begin{center}
\includegraphics[width=.8\textwidth]{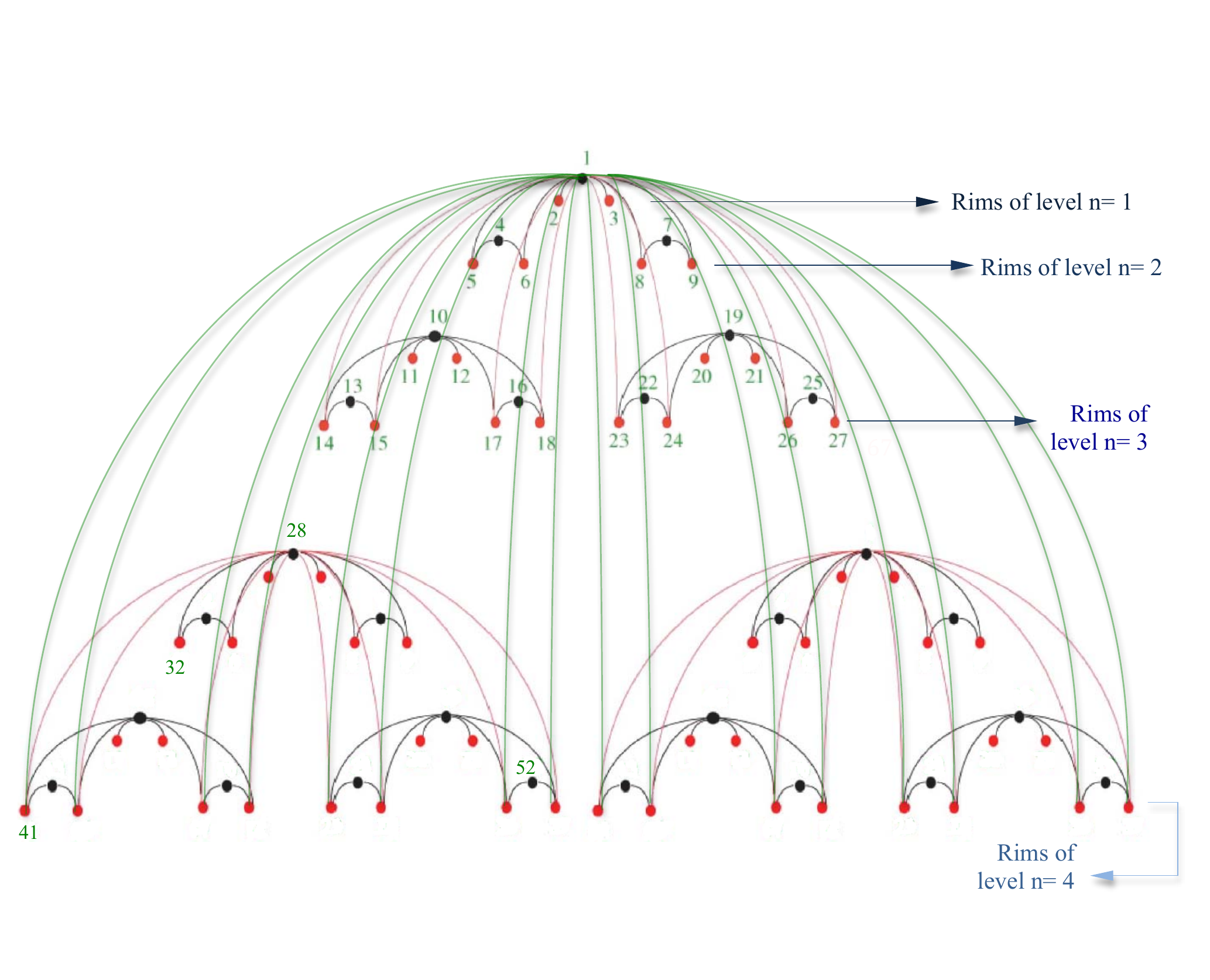}
\caption{\label{fig:grafo} (Color on line) The graph $\mathcal{G}_g$ of generation $g=4$. Darker nodes are hubs and brighter nodes are rims (of different subgraphs/levels). The labeling is complete for the subgraph in the top, and proceed analogously for the other subgraphs for which only a few labels have been inserted as example. Here $1$ represents the main hub, i.e. $H \equiv h_{4,4}$, nodes $\{ 2,3\}$ are rims of level $1$, nodes $\{ 5,6, 8, 9 \}$ are rims of level $2$ and so on, as specified. Also, focusing on node $14$, we notice that $10$ and $13$ are $h_{2,3}$ and $h_{1,3}$, respectively, while node $15$ is $r_{1,3}$, $\{17,18\} = \{ r_{2,3} \}$ and $\{23, 24, 26, 27\} = \{ r_{3,3} \}$.}
\end{center}
\end{figure*}

Now, in order to fix ideas we focus on a particular example of hierarchical network, introduced in \cite{Barabasi:2001PA} and further investigated in  \cite{Iguchi:2005PRE,Agliari:2009PRE,Manzotti:2010PRE,Zhang:2009JStat}, (see Fig.~\ref{fig:grafo}); other examples can be found in \cite{Ravasz:2003PRE,Zhang:2009PRE,Zhang:2010JStat,Zhang:2007EPJB,Nacher:2005PRE}. 

By denoting as $\mathcal{G}_g$ the graph of generation $g$, we have that $\mathcal{G}_0$ is given by a single node, also called ``root'', while $\mathcal{G}_1$ is a chain of length three obtained from $\mathcal{G}_0$ by adding two more nodes and connecting each of them to the root; the two nodes added are called ``rims'' (of level $1$).  
Similarly, at the second iteration, one introduces two copies of $\mathcal{G}_1$, whose rims are directly connected to the root: now the root is connected to the original two rims of level $1$ and to four rims of level $2$. 

Proceeding analogously, at the $g$-th iteration one
introduces two replica of the existing graph, i.e. of $\mathcal{G}_{g-1}$, and connects the root with all the new $2 \times 2^{g-1}$ rims, referred to as rims of level $g$.
Hence, the root turns out to be a hub connected with $2^n$ rims of level $n$, where $n \in [1,g]$, in such a way that its coordination number is $z_g=2(2^{g}-1)$, on the other hand, rims of level $n$ have a coordination number equal to $n$.

Given a rim of level $n$, here referred to as $r_n$, one can see that it is not only connected to the root but also to other ``minor hubs'' $h_{k,n}$, namely nodes that work as main hub for any subgraph $\mathcal{G}_k$, $k=1,...,n$, containing both $r_n$ and $h_{k,n}$; more precisely, we refer to $h_{k,n}$ as the hub of height $k$, with respect to a rim of level $n$ (see Fig.~\ref{fig:grafo}). 
The root will be also referred to as the main hub and denoted as $H \equiv h_{g,g}$. Also, given a node $i$ which is a rim of level $n$, we say that the set of rims of the same level and sharing with $i$ the same hub of height $k$ are rims shifted by $k$ with respect to $i$; this set is denoted as $\{ r_{k,n} \}$ and its cardinality is $|\{ r_{k,n} \}| = 2^n$ (see Fig.~\ref{fig:grafo}).
 The total number of nodes making up $\mathcal{G}_g$ is $N_g=3^g$, while the total number of rims is $\sum_{l=1}^g 2^n = 2(2^{g}-1) = z_g$.

Furthermore, we mention that the degree distribution for hubs is given by the
power law $P(k) \sim k^{- \gamma}$, with exponent $\gamma = \log 3 / \log 2 \approx 1.59$, while the remaining nodes follow an exponential degree distribution $P(k) \sim (2/3)^k$. For further details about the topological properties of $\mathcal{G}_g$ we refer to \cite{Barabasi:2001PA,Iguchi:2005PRE,Agliari:2009PRE,Manzotti:2010PRE,Zhang:2009JStat}.

We also notice that each subgraph making up the whole graph can be looked at as a module; connections between a module and the remaining graph are few (with respect to the size of the subgraph itself) and concern only the pertaining rims. Indeed, it is possible to determine a hierarchy of nodes, based on their degree of clustering,
consistently with \cite{Ravasz:2003PRE}: Although for $\mathcal{G}_g$ it is not possible to establish a one-to-one correspondence
between the clustering coefficient of a vertex and its degree, one can see that the clustering coefficient \footnote{For the graph under study the clustering coefficient should be defined according to the number of squares rather than triangles since the graph is devoid of triangles.} decreases with the degree.

The MFPT's on $\mathcal{G}_g$ have already been analyzed for special target locations \cite{Agliari:2009PRE,Manzotti:2010PRE,Zhang:2009JStat} and, before proceeding, it is worth recalling some results which may be useful in the following. 
In particular, for a simple RW on $\mathcal{G}_g$, the mean time to first reach the main hub $H$ starting from an arbitrary rim of level $g$ is \cite{Agliari:2009PRE}
\begin{equation}\label{eq:R_g}
T_g(H,r_g)= \frac{8}{3} \left( \frac{3}{2} \right)^g - 3, 
\end{equation}
while the mean time to first reach any of the $2^g$ rims of level $g$ starting from the main hub is
\begin{equation}\label{eq:H_g}
T_g(\{r_g\}, H) = \frac{4}{3} \left( \frac{3}{2} \right)^g - 1, 
\end{equation}
where the mean is taken over all possible paths; notice that the asymptotic behavior of $T_g(H,r_g)$ and $T_g(\{r_g\}, H)$ is the same, namely $\sim (3/2)^g$, even if the number of targets is $1$ and $2^g$, respectively.

\subsection{Labeling code and time to main hub}
We now introduce a method which allows to calculate straightforwardly the mean time to first reach the main hub starting from an arbitrary node. First of all, we need a proper labeling for nodes which exploits the topological features of the structure. Basically, we associate to  an arbitrary node $i$ belonging to the graph $\mathcal{G}_g$ a code, e.g. $\xi_i=(lrt..rrt)$, made up of $g$ letters properly chosen in the alphabet $\{t,r,l\}$, as we are going to explain. The whole graph can be looked at as the combination of three graphs of the previous generation: $\mathcal{G}_{g-1}$ (corresponding to $t$) and two copies of $\mathcal{G}_{g-1}$ arranged on the right ($r$) and on the left ($l$), respectively. Now, according to which of these main subgraphs $i$ belongs to, we have that $\xi_i^{1}$ is equal to either $t$, $r$ or $l$. Once detected the main subgraph, one proceeds analogously distinguishing the three subgraphs of second order, i.e. $\mathcal{G}_{g-2}$, and evaluating which contains the node $i$, hence determining $\xi_i^{2}$. Finally, at the $g$-th iteration, one is left with the subgraph $\mathcal{G}_1$, in such a way that its subgraphs are simply three nodes, one of them corresponds to $i$. 
For instance, referring to Fig.~\ref{fig:grafo}, we have: 
\begin{eqnarray}\label{eq:ex_times}
\nonumber
\xi_4&=&(ttlt),\\
\nonumber
\xi_{28}&=&(lttt),\\
\nonumber
\xi_{41}&=&(llll),\\
\nonumber
\xi_{52}&=&(lrrt).
\end{eqnarray}
As anticipated, we focus on arrangements where source and target belong to different modules; as we will see, this typically requires the passage through $H$, in such a way that we first need to calculate the time $T(H,i)$. For this aim, our addressing, while able to determine univocally a node, is somehow redundant, since, due to the intrinsic symmetry, the distinction between left and right subgraphs is unnecessary. For this reason, one can denote any of the two subgraphs in the bottom as $b$ in such a way that, for a graph of generation $g$, one could write
$\xi_i= (t^{i_1}b^{i_2}...t^{i_{k-1}}b^{i_k})$, with $\sum_{l=1}^k i_l= g$; also notice that, without loss of generality, we can always assume $\xi_i^{1} = t$ and $\xi_i^{k} =b$, with $i_1$ and $i_k$ possibly zero, while $i_l>0$ for $l \in [2,k-1]$.
Reading this string from right to left, we can write a general expression for the MFPT from $i$ to $H$. In fact, assuming that $i$ is a rim ($\xi_i^{k} =b$) of a certain inner subgraph $\mathcal{G}_{g_1}$, in order to reach $H$, we need to pass through the main hub of $\mathcal{G}_{g_1}$ itself, where $g_1$ is simply $i_k$. Now, the main hub of $\mathcal{G}_{g_1}$ is either $H$ (when $i_k=g$ or when $k=2$) or the main hub of a certain inner subgraph $\mathcal{G}_{g_2}$, where $g_2$ turns out to be $i_k + i_{k-1}$.
One can proceed analogously, bouncing from hub to rim and from rim to hub of larger and larger subgraphs, in such a way that the following general expression for the MFPT from $i$ to $H$ holds
\begin{eqnarray}
T(H,i) &\equiv& T(H, \xi_i) =   \\
\nonumber
&=& \sum_{l=1}^k [T_{j_l}(H,r_{j_l}) + T_{j_{l+1}}(\{r_{j_{l+1}}\},H) ],\\
\nonumber
j_l &=& g - \sum_{l'=1}^l i_{l'},
\end{eqnarray}
where we used $H$ to indicate the main hub of the (sub)graph considered (denoted by the index $j_l$) and $T_0=0$.
Recalling the examples above and using Eqs.~\ref{eq:R_g} and \ref{eq:H_g}, we write
\begin{eqnarray}\label{eq:ex_times2}
\nonumber
T(H,\xi_4) = T(t^2bt) &=& T_1(\{ r_1 \},H)+T_2(H,r_2),\\
\nonumber
T(H,\xi_{28}) =  T(bt^3) &=& T_3(\{r_3\},H) + T_4(H,r_4),\\
\nonumber
T(H,\xi_{41}) =  T(b^4) &=& T_4(H,r_4),\\
\nonumber
T(H,\xi_{52}) = T(b^3t) &=& T_1(\{r_1\},H)+T_4(H,r_4).
\end{eqnarray}
Of course, summing up such times over the whole set of nodes, one recovers the global mean first passage time $\tau_g \equiv \sum_{i \neq H} T(H,i)/(N-1)$ \cite{Agliari:2009PRE,Zhang:2009PRE}.

Now, in order to complete the calculation for the MFPT from an arbitrary source $i$ to a target $j$, we need $T(j,H)$, which can be calculated exploiting the centrality of $H$ and the self-similarity of the graph, by implementing a set of recursive equations. In order to preserve the generality of the method we focus on a particular class of targets, easily identifiable in generic hierarchical graphs, namely on rims of an arbitrary level $n$.

\subsection{MFPT's from hubs}
Beyond those discussed before, in order to calculate the MFPT from $H$ to a rim $r_n$, we need further quantities. First, let us introduce the following: $T_g(r_n,H)$, which represents the mean time to go from the main hub  to a rim of level $n$, $T_g(r_n,h_{k,n})$, which represents the mean time to go to a rim of level $n$ from a hub of height $k$ with respect to $r_n$, and $T_g(r_{n},r_{k,n})$, which represents the mean time to reach a rim of level $n$ from a rim of the same level, but ``shifted'' by $k$ (see Fig.~\ref{fig:grafo}) \footnote{In order to shrink the notation, in the expression $T_m(r_n,H)$ or $T_m(H,r_n)$, the point referred to as $H$ represents the main hub of the (sub)graph of generation $m$, whatever $n \leq m$.}. 

Then, we can write the set of equations:
\begin{eqnarray} \label{eq:r_H_impl}
T_g(r_n,H) &=& \frac{1}{z_g} + \frac{1}{z_g} \sum_{l=0}^{n-1} 2^l \left[ 1 + T_g(r_n, r_{l+1,n}) \right] \\
\nonumber
&+&   \frac{1}{z_g} \sum_{l=1}^{n-1}2^l \left[1 + T_l(H,r_l) + T_g(r_n,H) \right]  \\
\nonumber
&+&  \frac{1}{z_g} \sum_{l=n+1}^g 2^l \left[ 1 + T_l(H,r_l) + T_g(r_n,H) \right],
\end{eqnarray}
where the first term in the r.h.s. accounts for a direct jump from the root to the target, the second one accounts for shifted rims of level $n$ itself and the remaining terms account for rims of all levels other than $n$; 
\begin{eqnarray} \label{eq:r_h_impl}
T_g(r_n,h_{k,n}) &=& \frac{1}{z_k} + \frac{1}{z_k} \sum_{l=0}^{k-1} 2^l \left[ 1 + T_g(r_n, r_{l+1,n}) \right]\\
\nonumber
& +&   \frac{1}{z_k} \sum_{l=1}^{k-1} 2^l \left[ 1 + T_{l}(H,r_{l} )+ T_g(r_n,h_{k,n})\right]  ,
\end{eqnarray}
similarly to the previous case;
\begin{eqnarray} \label{eq:r_r_impl}
T_g(r_n, r_{k,n}) &=&   \frac{1}{n} \sum_{l=k}^n \left[ 1 + T_g(r_n, h_{l,n}) \right]\\
\nonumber
& +&  \frac{1}{n} \sum_{l=0}^{k-1} \left[ 1 + T_g(r_n, r_{k,n}  ) + T_l(\{r_{l} \},H) \right], 
\end{eqnarray}
where the first term in the r.h.s. accounts for the fact that to reach $r_n$ you need to pass through a common minor hub, or, possibly $H$ itself, while the second term accounts for bounces from the starting point to close (non common) minor hubs.

This system of recurrent equations can be solved by first focusing on Eqs.~\ref{eq:r_h_impl}-\ref{eq:r_r_impl} and building up the differences between terms for $k+1$ and $k$ so to get rid of the sums. The solutions found for $T_g(r_n,h_{k,n})$ and $T_g(r_n, r_{k,n} )$ are then plugged into Eq.~\ref{eq:r_H_impl} and, exploiting Eqs.~\ref{eq:R_g}-\ref{eq:H_g}, as well as proper initial conditions (e.g. $2 T(r_n,h_{1,n}) = 2 + T(r_n,  r_{1,n})$ and $T_g(r_{n,n}, r_n) = T_g(H,r_n) + T_g(r_n,H)$), one obtains closed form expressions which read as:
\begin{eqnarray} \label{eq:r_H}
\nonumber
T_g(r_n,H) &=&  \frac{c_n}{2} \left [ 1- \frac{1}{2^{n}} + \frac{n}{2^{n}} \psi(n)  \right] + 3\\ 
&-&4 \left( \frac{3}{2}\right)^{n-1} + 2^{2-n} (3^g - 2^g)\\
\label{eq:r_h}
T_g(r_n,h_{k,n}) &=& 2 \left( \frac{3}{2} \right)^k -1 + \frac{c_n}{2} \left[  1+ \frac{n}{2}  \phi(k,n) \right],\\
\label{eq:r_r}
T_g(r_n,r_{k,n}) &=& \frac{c_n}{2} \left[ 1 + \frac{1}{2^k} \frac{n}{n-k} + \frac{n}{2} \phi(k,n) \right],
\end{eqnarray}
where 
\begin{equation} 
\nonumber
c_n = 8 (3^g - 2^g)  [2n-1 - 2^{n-1} n\, \phi(n-1,n) + n\,  \psi(n)]^{-1},
\end{equation}
and
%
\begin{eqnarray}
\nonumber
\phi(k,n)= \sum_{i=0}^{k-1}\frac{2^{-i}}{n-1-i},\\
\nonumber
\psi(n)= \sum_{l=0}^{n-1} \left( \frac{1}{n-l} + 2^{l-1} \phi(l,n) \right).
\end{eqnarray}

All these formula have been successfully checked versus numerical estimates obtained by means of the pseudo Laplacian \cite{Fouss:2007IEEE}.

It is convenient to report the asymptotic ($N \rightarrow \infty$, i.e. $g$ large) expressions of previous quantities, which turn out to be the same for all of them, namely:
\begin{equation}
T_g(r_n,H) \sim T_g(r_n,h_{k,n}) \sim T_g(r_n, r_{k,n}) \sim \frac{3^g}{n},
\end{equation}
so that it is easy to see that the level $n$ plays algebraically: although the distance hub-rim and rim-rim is equal to $1$ and $2$, respectively, whatever $n, k$, rims added at larger generations are ``easier'' to be reached.
We also notice that the height of the minor hub considered or the shift among rims, just provide minor order corrections. 
In particular, once generation and level are fixed, the MFPT decreases with $k$, the reason is that, although the distance between starting point and target remains the same, independently of $k$, a large $k$ implies the passage through more connected hubs which are easier to be reached.

Analogous recursive equations can be implemented for the case of multiple targets (see for examples \cite{Agliari:2009PRE,Manzotti:2010PRE}), while here we just focus on the case of single target.

\subsection{Examples}
The results explained in the previous section, together with those summarized in the Sec.VA, allow to get an exact expression for the MFPT between two nodes $i$ and $j$, such that the path has to include a hub $h$. In this way one first detects the hub $h$ and then calculate $T(i,j)$ as a sum of the partial MFPT from $i$ to $h$ and from $h$ to $j$.
In order to clarify the procedure, we now present some examples where several kinds of situations are considered.
\bigskip\\
\emph{Source and Targets are both rims}\\
Being $i$ a rim of level $n$ and $j$ a rim of level $m$ with $n \neq m$ \footnote{When $i$ and $j$ share the same level, the intermediate hub is not univocally defined and these arguments do not apply.}, it is easy to see that, in order to go from $i$ to $j$ (or vice versa), one has to pass through $H$, so that 
\begin{eqnarray}
\label{eq:es2_1}
\nonumber
T(j,i) &=& T(r_m,r_n) = T_g(H,r_n)+T_g(r_m,H) \\
\nonumber
&=& T_n(H,r_n) + T_g(r_m,H) = \frac{8}{3} \left( \frac{3}{2} \right)^n - 3  \\
\nonumber
&+& \frac{c_m}{2} \left [ 1- \frac{1}{2^{m}} + \frac{m}{2^{m}} \psi(m)  \right] + 3 \\
\label{eq:es2}
&-&4 \left( \frac{3}{2}\right)^{m-1} + 2^{2-m} (3^g - 2^g) \sim 
\frac{3^g}{m}.
\end{eqnarray}
Hence, the leading term is typically $T_g(r_m,H)$. This also implies that the MFPT for the same nodes, but opposite direction, i.e. $T(r_n,r_m)$, differs from $T(r_m,r_n)$ and their ratio goes like $n/m$.

For example, let us refer to Fig.~\ref{fig:grafo} and let us fix $i=14$ and $j=41$ (or, of course, equivalent nodes). Then, we find
\begin{equation}
\nonumber
T_4(41,14)=T_3(H,r_3) + T_4(r_4,H)  = 6 + 1109/12 \approx 98.42,
\end{equation}
where we used Eqs.~\ref{eq:R_g} and \ref{eq:r_H}. 
Analogously,
\begin{equation}
\nonumber
T_4(14,41)=T_4(H,r_4) + T_4(r_3,H)  = 21/2 + 809/6 \approx 145.33.
\end{equation}
As expected from Eq.~\ref{eq:es2}, $T(14,41) > T(41, 14)$ due to the fact that from the main hub it is easier to reach a rim which belongs to higher levels.
\bigskip\\
\emph{Source and Targets are a hub and a rim}\\
Being $i$ a hub of height $k$ with respect to a set of rims of level $n$ and $j$ a rim of level $m$ with $n \neq m$, again, in order to go from $i$ to $j$ (or vice versa), one has to pass through $H$, so that
\begin{eqnarray}
\nonumber
T(j,i) &=& T_g(\{r_n\}, h_{k,n})+T_g(H,r_n) + T_g(r_m,H) \\
&=& T_k(\{r_k\},H)+T_g(H,r_n) + T_g(r_m,H) \\
\nonumber
&=&\frac{4}{3} \left( \frac{3}{2} \right)^k - 1 +  \frac{8}{3} \left( \frac{3}{2} \right)^n - 3 + \frac{c_m}{2} \\
\nonumber
&\times& \left [ 1- \frac{1}{2^{m}} + \frac{m}{2^{m}} \psi(m)  \right] + 3 -4 \left( \frac{3}{2}\right)^{m-1}\\
\label{eq:es1_1}
&+& 2^{2-m} (3^g - 2^g) \sim  \frac{3^g}{m}.
\end{eqnarray}
Hence, the leading term is typically $T_g(r_m,H)$. This is rather intuitive as $T_k(\{r_k\},H)$ accounts for trapping on a set of $2^k$ nodes, $T_g(H,r_n)$ for trapping on the main hub, while $T_g(r_m,H)$ for trapping on a single node with relatively small (equal to $m$) coordination number.

As an example, let us consider the graph $\mathcal{G}_4$ in Fig.~\ref{fig:grafo} and calculate
\begin{equation}
\nonumber
T_4(14,28)=\frac{7}{2} + \frac{21}{2} + \frac{809}{6} \approx 148.83.
\end{equation}
For comparison, let us also consider $\mathcal{G}_3$ for a ``rescaled case'', that is
\begin{equation}
\nonumber
T_3(5,10) =  2 + 6 + 54 = 62.
\end{equation}
\bigskip\\
\emph{Source and Targets are rims of the same subgraph}\\ 
Let us consider the subgraph $\mathcal{G}_{g'}$ of $\mathcal{G}_{g}$, where $g' < g$ and let us fix $i$ and $j$ as rims of level $m<n<g'$ of the subgraph, whose main hub has to be crossed in order to go from $i$ to $j$; for simplicity, let us assume that $j$ also belongs to the set $\{ r_g\}$ in $\mathcal{G}_{g}$.
Therefore we can write 
\begin{eqnarray}
T(j,i) &=& T_{g'}(H,r_{m}) + T_g(r_n,h_{g',n})\\
&=& T_{m}(H,r_{m}) + T_g(r_n,h_{g',n})\\
\nonumber
&=& \frac{8}{3} \left( \frac{3}{2}\right)^m  + 2 \left( \frac{3}{2} \right)^{g'} - 4 + \frac{c_n}{2} \left[  1 + \frac{n}{2} \phi(k,n) \right] \\
\nonumber
&\sim& \frac{3^g}{n}.
\end{eqnarray}

For example, still referring to the labeling of Fig.~\ref{fig:grafo} let us fix $i=32$ and $j=41$. Then, for $g=4$ we find
\begin{equation}
\nonumber
T(41,32)=T_2(H,r_2) + T_4(r_4,h_{3,4})  = 3 + 887/12 \approx 76.92,
\end{equation}
where we used Eqs.~\ref{eq:R_g} and \ref{eq:r_h}.
Analogously, for $g=3$ we find
\begin{equation}
\nonumber
T(14,11)=T_1(H,r_1) + T_3(r_3,h_{2,3})  = 1 + 211/6 \approx 38.17.
\end{equation}

\section{Conclusions}
In this work we introduced general methods to calculate exactly first-passage quantities on self-similar networks, which are defined recursively.
In particular, we focused on the mean first-passage time from a source $S$ to one or to several target sites $\{ T_i \}$, and on the splitting probability, namely the probability to reach, starting from $S$, a given target before all other targets. 
Indeed, these quantities allow a sound description of a wide range of dynamical processes such as diffusion limited reactions or  search processes embedded in complex media \cite{Benichou:2011b,Obenichou:2008}.

In general, the methods introduced strongly rely on the recursivity of the underlying structure, namely on the fact that the whole graph can be built according to a recursive procedure: at the $n$-th iteration the graph $\mathcal{G}_n$ is obtained by properly combining a finite number of graphs $\mathcal{G}_{n-1}$, each corresponding to the graph itself at the previous iteration.

Recursive networks where replicas meet at a single node are amenable to exact analysis by renormalization techniques and we show that the above-mentioned first-passage quantities can be recast as solutions of simple matricial equations. 
We also consider examples where replicas are connected by links and exact decimation is no longer accomplishable; then, one can impose a number of coupled equations, each corresponding to a set of equivalent nodes.

Hence, a broad range of topologies can be addressed via a unifying approach: we considered explicitly recursive networks as diverse as finitely ramified fractals (Sierpinski gasket, T-fractal), scale-free (trans)fractals ($(u,v)$-flowers), non-fractals, mixtures between fractals and non-fractals (Song-Havlin-Makse networks), non-decimable hierarchical graphs (Barab\'{a}si-Ravasz-Vicsek network).

In any case, calculations performed are exact; results previously obtained for special cases of target arrangements are recovered and extended to account for more general configurations.

\appendix
\section{The T-graph}

In this section we apply  the  method that has been presented  in section \ref{general}   to the T-graph (see Fig. \ref{fig:ex_fractals}). We first give  the matrices $\mathcal{M}_{i_k}$, the vectors $\mathcal{V}_{i_k}$ and the exit time $\tau_k$. It is then possible to calculate explicitly splitting probabilities and MFPTs for any starting point on the network. For example, we calculate explicitly two different types of averages: the MFPT averaged over starting points of a given level  and the MFPT averaged over all starting sites with a given connectivity. In the particular case where the average is performed over all starting positions, we recover the result of  Ref \cite{Agliari:2008PRE}.

\begin{figure}
\begin{center}
\includegraphics[width=9cm]{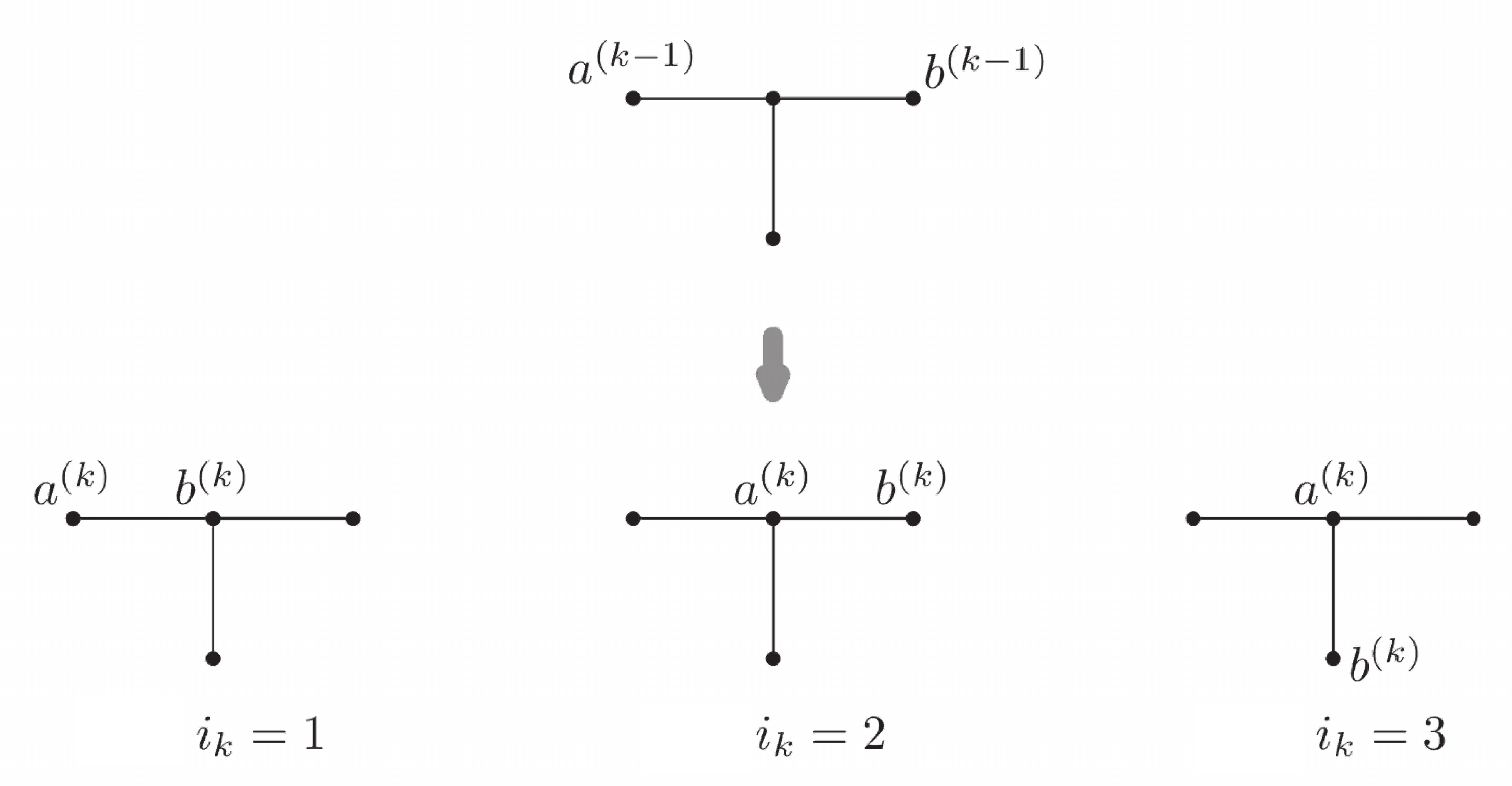}
\caption{\small Renormalization scheme and indexes for the T-graph}
\label{fig:renorm_arbreT}
\end{center}
\end{figure}

\subsection{General results}

Quantities $\mathcal{P}$ and $\mathcal{T}$ are here 2-dimensional vectors, and will be denoted $\begin{pmatrix} a\\b\end{pmatrix}^{(k)} $. Figure \ref{fig:renorm_arbreT} shows the chosen (but not unique) labeling scheme ($p=3$).

\paragraph*{Matrices and vectors.}  One finds
\begin{eqnarray}
{\cal M}_1 = \begin{pmatrix} 1&0\\1/2&1/2 \end{pmatrix} , && {\cal M}_2 = \begin{pmatrix} 1/2&1/2\\0&1 \end{pmatrix} , \nonumber \\
 {\cal M}_3 &=& \begin{pmatrix} 1/2&1/2\\1/2&1/2 \end{pmatrix}.
\end{eqnarray}
 \be
 {\cal V}_1 = \begin{pmatrix} 0\\2 \end{pmatrix}, ~~{\cal V}_2 = \begin{pmatrix} 2\\0 \end{pmatrix},~~ {\cal V}_3 = \begin{pmatrix} 2\\3 \end{pmatrix}.
 \fin

\paragraph*{Crossing time.} One finds
\be
 \tau_k =  6^{n-k}.
 \fin

\subsection{Examples of averaged  MFPTs} 

Let us consider the case where the target is located at the extreme left vertex of the T-graph : $T = a^{(0)}$. Note that an average MFPT calculated for that target for a network of generation $n $ is equal to an average for the next ($n+1$) generation with a target at the center.   
Using that setting, the equivalent of equation (\ref{eq:sierp_mfpt_nonrecursive}) is:
\be
\mathcal{T}^{(k)} = \sum_{l=0}^k \tau_l \left( {\cal M}_{i_k} \cdots {\cal M}_{i_{l+1}} \right) {\cal V}_{i_{l}}    ~~{\rm with}~ {\cal V}_{i_0} = \begin{pmatrix} 0\\1 \end{pmatrix}.
\fin

\paragraph*{MFPT averaged over the sources of level $k$. }
Following the  method developed  in section \ref{sec:mfpt_avg}, we make use of Eq. (\ref{eq:mfpt_avg_general}) and obtain:
\begin{eqnarray}\label{eq:tgraph_samelevel_sum}
\sum_{paths} \begin{pmatrix} a\\b\end{pmatrix}^{(k)}& = &6^{n-1} \left[ 4-2^{-k} \right] \begin{pmatrix} -1\\1 \end{pmatrix}  \nonumber \\
&+& \frac{6^n}{5}  \left[ 4\times3^k - 3\times2^{-(k+1)} \right] \begin{pmatrix} 1\\1 \end{pmatrix} .
\end{eqnarray}

As a result of the indexation scheme that has been chosen, $S^{(k)} = \sum_{paths} \, b^{(k)}\,$; in other words the MFPT summed over all sources of level $k$ is equal to the 2nd coordinate of Eq. (\ref{eq:tgraph_samelevel_sum}). Therefore, using the fact that there are $3^k$ sources, one gets :
\be\label{eq:tgraph_avg}
{\overline{\aver{T}}}^{(k)} = \frac{4}{5}\, 6^n \left[ 1+\frac{5}{6\times3^k}-\frac{7}{2\times6^{k+1}} \right].
\fin
For the particular case $k=n$, the latter expression gives the exact result derived in \cite{Agliari:2008PRE}.

\paragraph*{MFPT averaged over starting points of given connectivity.} 
 The nodes of the T-graph have connectivity $\kappa = 1 $ or 3. It is possible to use (\ref{eq:tgraph_samelevel_sum}) to determine the MFPT averaged over all the $N_{\kappa=3}$ starting sites of connectivity $\kappa=3$ (which of course gives access to the similar quantity for the starting sites of connectivity 1). Indeed, an examination of the labeling scheme applied to the entire network shows that the first coordinate of (\ref{eq:tgraph_samelevel_sum}) is equal to $\displaystyle 2\, N_{\kappa=3}\, {\overline{\aver{T}}}_{\kappa = 3}^{(k)}$. Then, using the fact that $\displaystyle N_{\kappa=3}=\frac{3^k-1}{2}$, one obtains : 
\be\nonumber
{\overline{\aver{T}}}_{\kappa=3}^{(k)} = \frac{4}{5}\, 6^n \left[ 1+\frac{1}{2\times3^{k+1}} \times \frac{1-2^{-k}}{1-3^{-k}} \right].
\fin

\section{The $(u,v)$-flowers}
\begin{figure}
\begin{center}
\includegraphics[height=8.5cm]{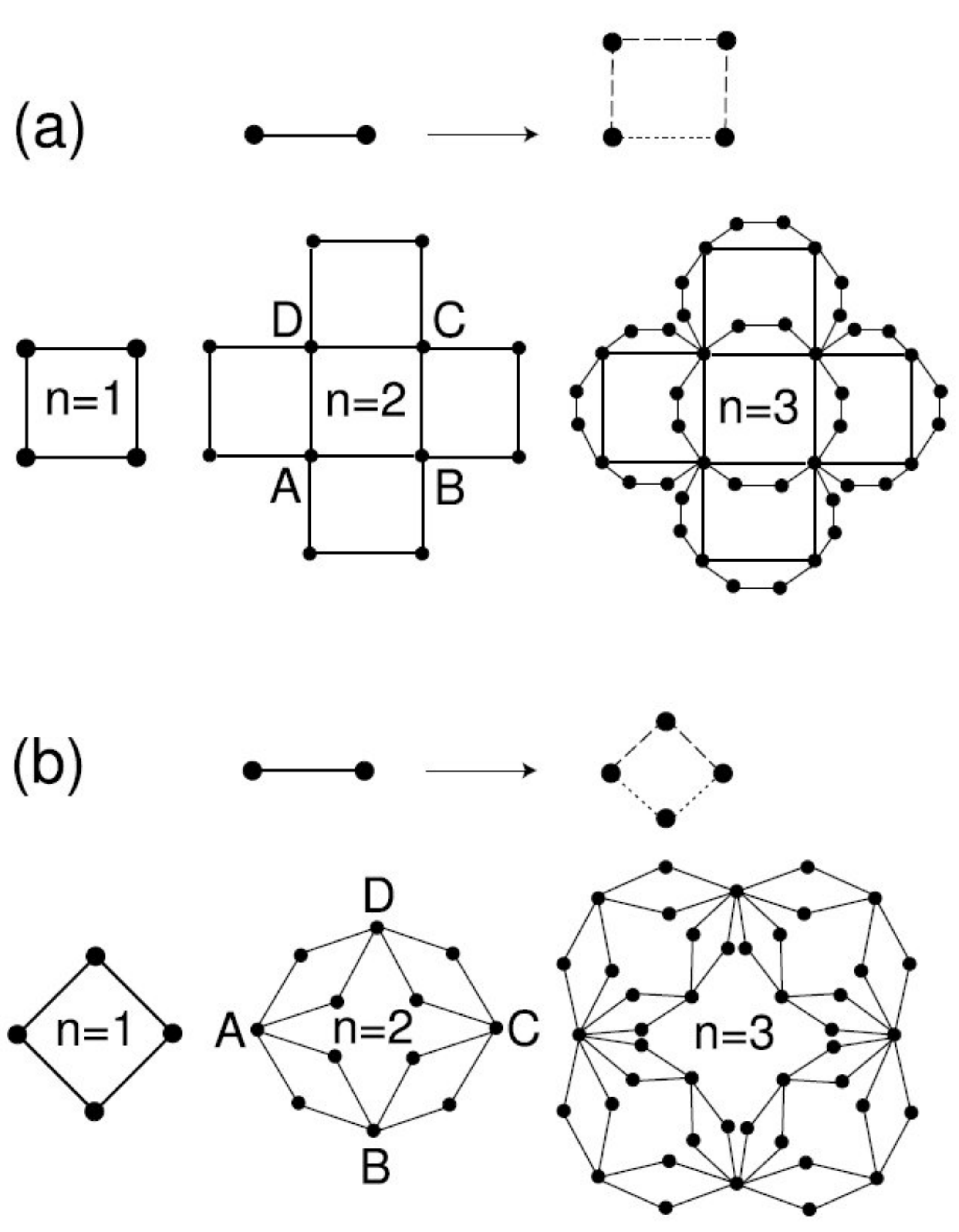}
\caption{\small Renormalization scheme for two $(u,v)$-flower networks : {\bf (a)} $u=1,v=3$ ; {\bf (b)} $u=2,v=2$. Figure reproduced from \cite{Rozenfeld:2007NJP}.}
\label{fig:renorm_flowers}
\end{center}
\end{figure}

In this section, we consider the case of the $(u,v)$-flower networks introduced in \cite{Rozenfeld:2007PRE} as examples of deterministic scale-free
networks, that are either fractal  or small-world depending on the values of the  two parameters
$u$ and $v$.  The algorithm to build the $(u,v)$-flower networks is as follows:  one starts (generation $n = 0$) with two sites connected by a link; generation $n + 1$ is obtained recursively by replacing each link by two parallel paths of respectively $u$ and $v$ links. In order to simplify some notations, we define $w \equiv u+v$, and take (without loss of generality) $u \le v$.  Examples of $(1,3)$ and $(2, 2)$-flowers are shown in Figure \ref{fig:renorm_flowers}. Let us remark that the connectivity of a site is determined by its level: $\kappa_k = 2^{n-k+1}$ when $k \ge 1$ and $\kappa_0 = 2^n$.

We focus on the examples $(u=1, v=3)$, $(u=2 , v=2)$ and $(u=2 , v=3)$. As will be shown in Section \ref{sec:flowers_generalities}, the $(1,3)$-flower is a transfractal (small-world) network, whereas the $(1,3)$ and $(2,3)$-flowers are fractal networks. Note that 
the GMFPT for a target on level 0 was previously obtained by Zhang et al. \cite{Zhang:2009fl} in the cases $(1,3)$ and $(2, 2)$ flowers. We show here that the MFPT can be obtained for any  $(u,v)$-flower network for any starting site, and that averages over different classes of starting sites can be obtained.


\begin{figure}
\begin{center}
\includegraphics[height=8.5cm]{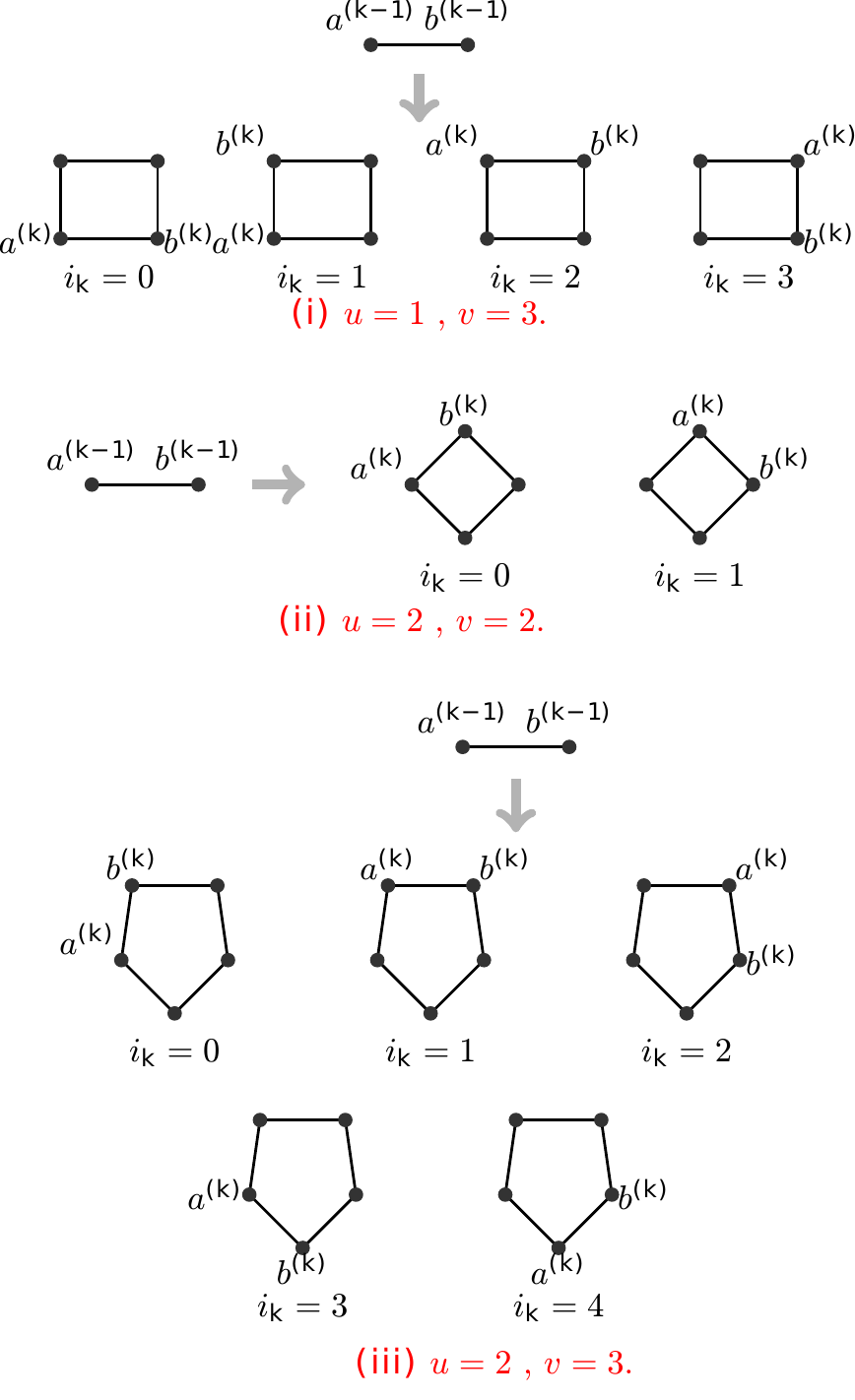}
\caption{\small Color online. $(u,v)$-flower networks : indexes.} 
\label{fig:flower_labels}
\end{center}
\end{figure}

\subsection{General properties of the $(u,v)$ flowers}\label{sec:flowers_generalities}


The number of links of  level $k$ is $w^k$ and the number of sites of level $k$ is given by \cite{Rozenfeld:2007PRE}:
\be\label{eq:flowers_vol}
N_k = \frac{w-2}{w-1}\, w^k +\frac{w}{w-1}.
\fin
The value of the  parameter $u$  splits the  scaling of the diameter $R$ with respect to $n$ into two distinct cases: 
\be\label{eq:flowers_diameter}
R \sim \left\{ \begin{array}{ll}
                  (v-1)\,n &~{\rm for}~ u = 1,\\
                  u^n &~{\rm for}~ u \ge 2.
                \end{array}
\right.
\fin
Therefore, when  $u=1$, the network is a small-world. When $u \ge 2$, it is a fractal network, and its fractal dimension is obtained by combining equations (\ref{eq:flowers_vol}) and (\ref{eq:flowers_diameter}):
\be\label{eq:flowers_fracdim}
d_f = \frac{\ln(u+v)}{\ln u}.
\fin

\paragraph*{Crossing time.} Following the method presented in Section \ref{sec:sierp_MFPT}, we find a general formula for $\tau_k$:
\be
\tau_k = \Bigg( \frac{u+v}{ 2- \frac{u-1}{u}-\frac{v-1}{v}}  \Bigg) ^{n-k} .
\fin
This result holds also for $u=1$.


\subsection{$(1,3)$-flower}

 We first study the case $u=1$, $v=3$, as an example of small world network. Figure \ref{fig:flower_labels} (i) shows the labeling scheme that has been chosen. We find that $\tau_k = 3^{n-k}$ and:
\begin{eqnarray}
 {\cal M}_0 = \begin{pmatrix} 1&0\\0&1 \end{pmatrix}, && {\cal M}_1 = \begin{pmatrix} 1&0\\2/3&1/3 \end{pmatrix}, \nonumber\\
 {\cal M}_2 = \begin{pmatrix} 2/3&1/3\\1/3&2/3\end{pmatrix} , && {\cal M}_3 = \begin{pmatrix} 1/3&2/3\\0&1\end{pmatrix}, \\
 {\cal V}_0 = \begin{pmatrix} 0\\0 \end{pmatrix} , ~{\cal V}_1 = \begin{pmatrix} 0\\2 \end{pmatrix}, && {\cal V}_2 = \begin{pmatrix} 2\\2 \end{pmatrix} , ~{\cal V}_3 = \begin{pmatrix} 2\\0 \end{pmatrix}.
 \end{eqnarray}
 It is then possible to calculate exactly the splitting probabilities and MFPTs for any starting site on the network.

\paragraph*{MFPT averaged  over starting sites of level $k$.} We assume that the target is located at  $a^{(0)}$ in level 0, and we  apply the  method developed  in Sections \ref{sec:mfpt_avg} (derivation of $\sum_{paths} \mathcal{T}^{(k)}$) and \ref{sec:song_avg_same_level} (substraction of branches repetitions).  We first use formula (\ref{eq:mfpt_avg_general}) with $ \mathcal{T}^{(0)} = \begin{pmatrix} 0\\3^n\end{pmatrix} $ to derive 
\be
  \sum_{paths} \mathcal{T}^{(k)} = 3^n \, \bigg[ 2^{k-1} \begin{pmatrix} -1\\1 \end{pmatrix} + 4^k \bigg(1-\frac{1}{2\times3^k} \bigg) \begin{pmatrix} 1\\1 \end{pmatrix} \bigg].
\fin
We then focus on the 2nd coordinate of the latter expression, but we  need to take care of the fact that branches $i_k = 0$ and $i_k = 3$ generate two contributions of the same source $b^{(k)}$: a way to avoid this repetition is to substract the same quantity for all the levels lower than $k$, just like in Eq. (\ref{eq:song_substracting_levels}) (with $x=1$).
We then get:
\be
 \left(N_k-1\right) {\overline{\aver{T}}}^{(k)} =   2\times3^{n-1}\,4^{k} +3^{n-k}\,4^k - 2\times3^{n-1}
\fin
with $N_k = \frac{2}{3}\left(4^k + 2\right)$.
For the particular case $k=n$, the latter expression is in agreement with \cite{Zhang:2009fl}.

\subsection{$(2,2)$-flower}

In this example we find $\tau_k =   4^{n-k}$ and:  
\begin{eqnarray}
{\cal M}_0 = \begin{pmatrix} 1&0\\1/2&1/2 \end{pmatrix}, && {\cal M}_1 = \begin{pmatrix} 1/2&1/2\\0&1\end{pmatrix}, \\
{\cal V}_0 = \begin{pmatrix} 0\\1 \end{pmatrix}, && {\cal V}_1 = \begin{pmatrix} 1\\0 \end{pmatrix}. 
\end{eqnarray}
 It is then possible to calculate exactly the splitting probabilities and MFPTs for any starting site on the network.

\paragraph*{MFPT averaged  over starting sites of level $k$.} Again, the target is located on $a^{(0)}$ and $ \mathcal{T}^{(0)} = \begin{pmatrix} 0\\4^n\end{pmatrix} $. The repeted  branches  must be taken into account: the method has been given in Section \ref{sec:song_avg_same_level}. Here $\nu_0 = \nu_1 = 2$. Using formula (\ref{eq:song_raw_sum_paths}) we get:
\be
  \sum_{paths} \mathcal{T}^{(k)} = 4^n \, \bigg[ 2^{k-1} \begin{pmatrix} -1\\1 \end{pmatrix} + \frac{1}{6} \bigg(4^{k+1}- 1 \bigg) \begin{pmatrix} 1\\1 \end{pmatrix} \bigg].
\fin
Then we focus on the 2nd coordinate of the latter expression, and apply equation (\ref{eq:song_substracting_levels}):
\be
 \left(N_k-1\right) {\overline{\aver{T}}}^{(k)} = \frac{4^n}{18} \big[ 2\times4^{k+1} +3k +10 \big]
\fin
with $N_k = \frac{2}{3} \left(4^k + 2\right)$.
For the particular case $k=n$, the latter expression is in agreement with  \cite{Zhang:2009fl}.


\subsection{$(2,3)$-flower}

The flower of parameters $u=2$, $v=3$, is a fractal network characterized by $\tau_k = 6^{n-k}$. Figure \ref{fig:flower_labels} (iii) shows the chosen labeling scheme. One finds 
\begin{eqnarray}
 {\cal M}_0 = \begin{pmatrix} 1&0\\2/3&1/3 \end{pmatrix}, &&  {\cal M}_1 = \begin{pmatrix} 2/3&1/3\\1/3&2/3 \end{pmatrix}, \nonumber\\ 
 {\cal M}_2 = \begin{pmatrix} 1/3&2/3\\0&1\end{pmatrix}, &&\\
   {\cal M}_3 = \begin{pmatrix} 1&0\\1/2&1/2\end{pmatrix},&&  {\cal M}_4 = \begin{pmatrix} 1/2&1/2\\0&1\end{pmatrix},\nonumber\\
 {\cal V}_0 = \begin{pmatrix} 0\\2 \end{pmatrix} , {\cal V}_1 = \begin{pmatrix} 2\\2 \end{pmatrix}, && {\cal V}_2 = \begin{pmatrix} 2\\0 \end{pmatrix} , \nonumber \\
 {\cal V}_3 = \begin{pmatrix} 0\\1 \end{pmatrix}, &&{\cal V}_4 = \begin{pmatrix} 1\\0 \end{pmatrix}.
 \end{eqnarray}
  It is then possible to calculate exactly the splitting probabilities and MFPTs for any starting site on the network.

\paragraph*{MFPT averaged  over starting sites of level $k$.} The target is located on  $a^{(0)}$. We again  apply  formula (\ref{eq:mfpt_avg_general}) with $ \mathcal{T}^{(0)} = \begin{pmatrix} 0\\6^n\end{pmatrix} $ and find : 
\be
  \sum_{paths} \mathcal{T}^{(k)} = 6^n \, \bigg[ 2^{k-1} \begin{pmatrix} -1\\1 \end{pmatrix} + 5^{k-1} \bigg(\frac{7}{2}-\frac{1}{6^k} \bigg) \begin{pmatrix} 1\\1 \end{pmatrix} \bigg].
\fin
Substracting the same quantity for all the levels lower than $k$ (see Eq. (\ref{eq:song_substracting_levels})) finally yields:
\be
 \left(N_k-1\right) {\overline{\aver{T}}}^{(k)} = 6^n \left[ \frac{21}{8}\,5^{k-1} - 7\times5^{k-1}\,6^{-k} +\frac{15}{8} \right]
\fin
with $N_k = \frac{3\times5^k + 5}{4}$.


\section*{Acknowledgement}
EA is grateful to the Italian Foundation ``Angelo della Riccia'' for financial support.
The research belongs to the strategy of exploration funded by the FIRB project RBFR08EKEV
which is acknowledged.


\end{document}